\documentclass[sigconf,nonacm]{acmart}

\usepackage{colortbl}
\usepackage{tikz}
\usepackage{enumitem}
\usepackage{verbatim}
\usepackage{float}
\usepackage{caption}
\usepackage{amsmath,amsthm,amsfonts,amssymb,amscd,bm,amsbsy}
\usepackage{balance}
\usepackage[TABBOTCAP]{subfigure}
\usepackage{enumerate}
\usepackage{fancyhdr}
\usepackage{pifont}
\usepackage{mathrsfs}
\usepackage{listings}
\usepackage{textcomp}
\usepackage{xcolor}
\usepackage{hyperref}
\usepackage{ragged2e}
\usepackage[utf8]{inputenc}
\usepackage{graphicx}
\usepackage{lipsum}
\usepackage{indentfirst}
\usepackage{cleveref}
\usepackage{tabularx}
\usepackage{balance}
\usepackage{wrapfig}
\usepackage{multirow}
\usepackage{capt-of}
\usepackage[many]{tcolorbox}
\usepackage{placeins}
\makeatletter 
\usepackage{algorithm}
\usepackage[noend]{algpseudocode}
\algrenewcommand\algorithmicthen{}
\algrenewcommand\algorithmicdo{}
\usetikzlibrary{arrows.meta,bending}
\renewcommand\footnotetextcopyrightpermission[1]{}

\newcommand{\eg}{\textit{e.g.}}

\definecolor{reviewA}{HTML}{FF0000}
\definecolor{reviewB}{HTML}{00FF00}
\definecolor{reviewC}{HTML}{0000FF}

\definecolor{gjr}{HTML}{FF0000}

\newcommand{\ourname}{Vertumnus}

\newtcolorbox{reviewercomment}{
    breakable,
    colback=gray!10,
    colframe=gray!40, 
    arc=3pt, 
    boxrule=1pt, 
    left=6pt, right=6pt, top=2pt, bottom=2pt,
}

\title{Adaptive Context Parallelism for Production LLM Serving}

\author{Jiarui Guo}
\affiliation{%
  \institution{Peking University}
  \city{}
  \country{}
}

\author{Rongle Wang}
\affiliation{%
  \institution{Peking University}
  \city{}
  \country{}
}

\author{Peijun Huang}
\affiliation{%
  \institution{Peking University}
  \city{}
  \country{}
}

\author{Zongwei Lv}
\affiliation{%
  \institution{Peking University}
  \city{}
  \country{}
}

\author{Ziqing Wang}
\affiliation{%
  \institution{Alibaba Group}
  \city{}
  \country{}
}

\author{Kan Liu}
\affiliation{%
  \institution{Alibaba Group}
  \city{}
  \country{}
}

\author{Tao Lan}
\affiliation{%
  \institution{Alibaba Group}
  \city{}
  \country{}
}

\author{Lin Qu}
\affiliation{%
  \institution{Alibaba Group}
  \city{}
  \country{}
}

\author{Xiaolin Wang}
\affiliation{%
  \institution{Peking University}
  \city{}
  \country{}
}

\author{Tong Yang}
\affiliation{%
  \institution{Peking University}
  \city{}
  \country{}
}

\renewcommand{\shortauthors}{Jiarui Guo et al.}
\acmConference[arXiv preprint]{arXiv preprint}{}{}
\begin{document}
\allowdisplaybreaks

\twocolumn 
\setlength{\subfigcapskip}{-0.15cm}
\setlength{\subfigbottomskip}{-0.05cm}

\begin{abstract}
As LLM context windows expand and input sequences grow longer, serving systems face increasing computational and memory demands.
Context parallelism (CP), which partitions the input sequence across multiple ranks to parallelize the computation, has therefore become increasingly important for efficient LLM serving.
However, existing CP-enabled systems either rely on static CP configurations or adjust the CP degree only for active requests or batches.
In this paper, we present \ourname{}, an adaptive CP serving system designed for heterogeneous and evolving workloads.
At the request level, \ourname{} routes requests among workers with different CP degrees using a placement cost that combines predicted queuing delay, cache-aware prefill time, and GPU-time cost.
At the cluster level, \ourname{} adapts the worker composition through seconds-scale split and merge operations as workload demand changes.
\ourname{} further introduces a global prefix-cache management policy that coordinates cache placement and replication among workers with the same or different CP degrees, preserving cache locality as request assignments and worker composition change.
Experiments on a 64-GPU cluster with public and production workloads show that, under the highest evaluated loads, \ourname{} reduces mean TTFT by up to 28.1\% and improves token-weighted SLO attainment by up to 13.3 percentage points over the strongest baseline.
\end{abstract}

\maketitle
\sloppy
\renewcommand{\thefootnote}{}
\footnotetext{Kan Liu (liukan.lk@alibaba-inc.com) and Tong Yang (yangtong@pku.edu.cn) are corresponding authors. }
\renewcommand{\thefootnote}{\arabic{footnote}}

\section{Introduction}
\label{sec:intro}

\begin{figure*}[!ht]
    \centering
    \begin{minipage}{.5\textwidth}{
    \centering
    \includegraphics[height=9em]{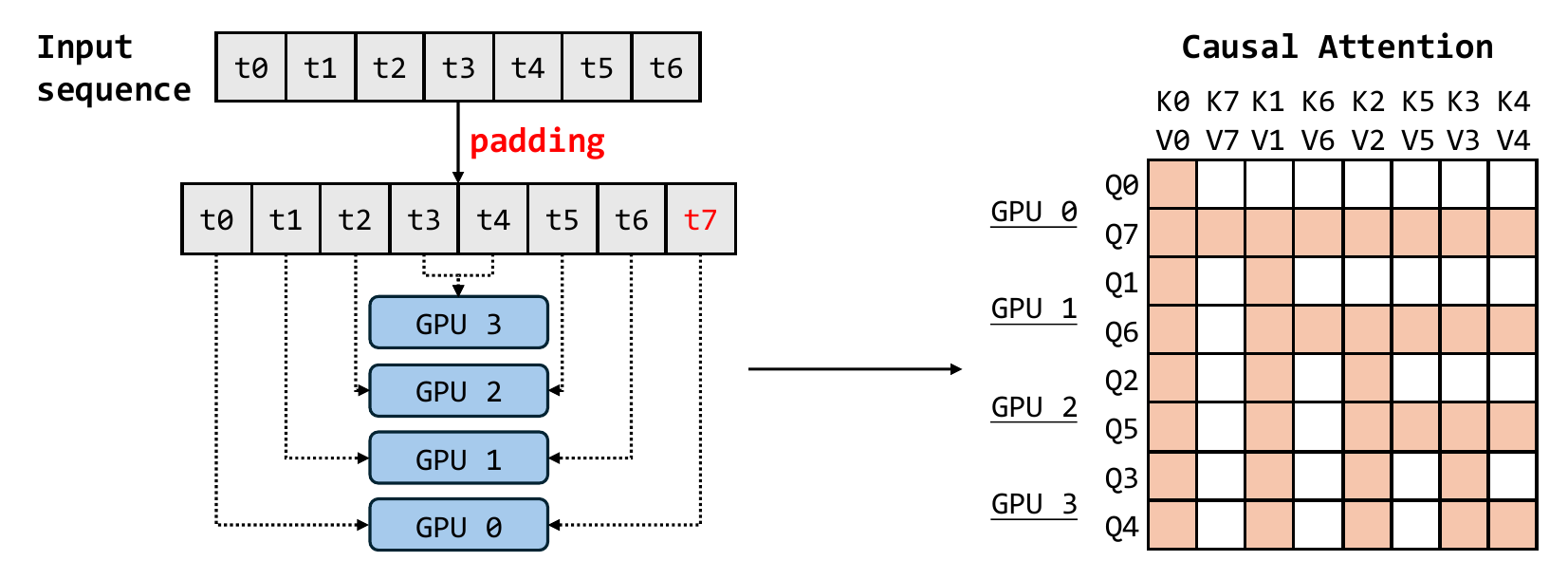}
    \vspace{-0.05in}
    \caption{An example of context parallelism when $C=4$.}
    \vspace{-0.2in}
    \label{fig:cp}}
    \end{minipage}%
    \begin{minipage}{.5\textwidth}{
    \centering
    \includegraphics[height=10em]{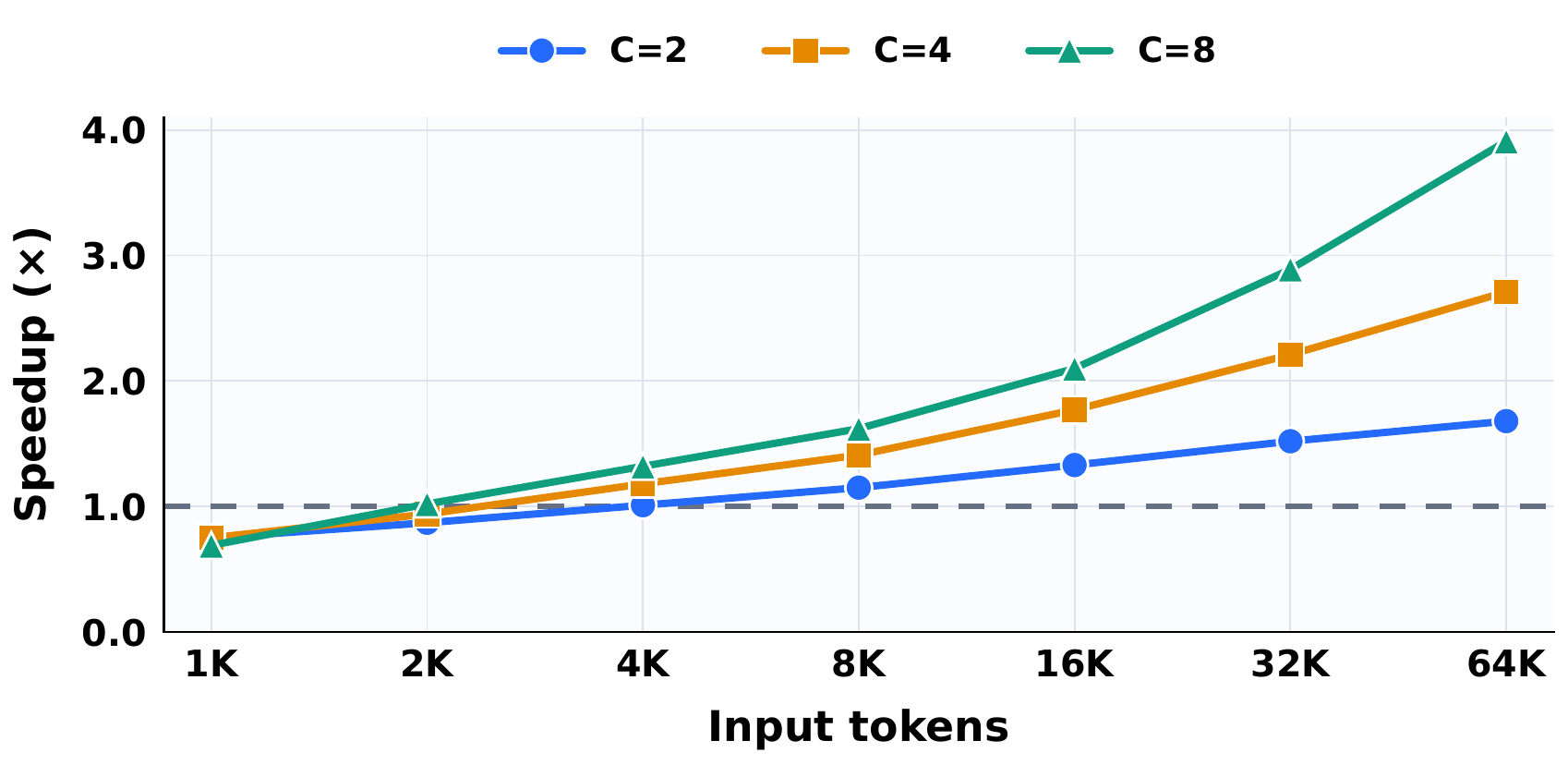}
    \vspace{-1em}
    \vspace{-0.05in}
    \caption{Effects of CP on inputs with different lengths.}
    \vspace{-0.2in}
    \label{fig:cp-effect}}
    \end{minipage}
\end{figure*}

Recent years have witnessed remarkable advances in large language models (LLMs), driven by rapid improvements in model capabilities, context length, and model scale \cite{xu2026deepseek, comanici2025gemini, touvron2023llama}.
LLMs are now powering a diverse range of applications, \eg{} conversational assistants \cite{achiam2023gpt}, retrieval-augmented generation \cite{lewis2020retrieval}, long-document understanding \cite{bai2025longbench}, and autonomous agents \cite{talebirad2023multi}. 
As LLMs become deeply integrated into products and enterprise workflows, LLM inference is increasingly delivered as a continuously available online service rather than executed only as isolated offline tasks \cite{sun2026hygen, jiang2025neo}, placing LLM serving on the critical path of production applications.

Therefore, efficient LLM serving systems have become a critical component of modern AI infrastructure \cite{kwon2023efficient, zheng2024sglang, tan2026rtp}. 
In production, the workloads handled by these systems exhibit three important characteristics. 
\ding{172} Request lengths are highly heterogeneous.
Conventional question-answering requests may contain only a few hundred tokens, whereas document analysis, retrieval-augmented generation, and complex agentic requests may contain tens of thousands of tokens or more \cite{wang2025llm}. 
\ding{173} Workloads are temporally dynamic.
Both the request arrival rate and the composition of short and long requests can change substantially over time \cite{wang2025burstgpt, mo2026serving}.
\ding{174} Prefix reuse is becoming increasingly prevalent.
Shared system prompts, multi-turn conversations, and agent/subagent workflows often produce long repeated prefixes, making prefix caching critical for avoiding redundant prefill computation \cite{pan2026kvflow, ye2024chunkattention, yang2026beluga}. 
Efficiently serving such workloads requires matching computational resources to both individual requests and evolving aggregate demand, while exploiting prefix reuse and maintaining overall GPU efficiency.

These requirements become particularly demanding for long-context requests.
As context lengths continue to grow, \textbf{context parallelism (CP)}\footnote{
Some early works use sequence parallelism (SP) to refer to attention-level sequence partitioning, which we call CP.
In this paper, CP partitions the input sequence and attention computation across ranks, whereas SP refers to partitioning the activations of non-matrix-multiplication operations, such as LayerNorm and dropout, along the sequence dimension.} has emerged as an important technique for accelerating long-context processing \cite{li2023sequence, liu2024ringattention, brandon2023striped}. 
In CP, the input tokens of a request are partitioned across multiple ranks, each mapped to one GPU, and the attention computation is distributed among these ranks.
This allows long requests to exploit the compute capacity of multiple GPUs and reduce per-GPU memory pressure, potentially reducing time-to-first-token (TTFT) at the cost of additional cross-GPU communication and a larger resource footprint.

Like workload intensity and prefix-cache locality, the CP degree is an important factor that a serving scheduler must consider.
Different CP degrees induce different request latency, communication overhead, GPU consumption, and aggregate serving capacity.
However, to the best of our knowledge, the current landscape leaves three important gaps.
\textit{First,} most existing LLM serving systems either lack support for CP or do not treat the CP degree as a first-class scheduling dimension \cite{zhang2026simple, yu2022orca, sun2024llumnix}.
Their scheduling policies typically optimize request characteristics, cache locality, or worker load without modeling how these factors interact with the CP degree.
\textit{Second,} systems that support CP commonly determine each worker's CP degree at deployment time \cite{yang2025context}.
When workers with multiple CP degrees coexist, GPUs are statically partitioned among persistent CP workers.
Since requests are usually routed using offline profiles and length-based rules, such static provisioning cannot adapt the worker composition to workload evolution.
Moreover, routing solely by input length may be suboptimal, as it ignores worker-specific prefix reuse and current load.
\textit{Third,} although recent systems dynamically adjust CP at request, batch, or iteration granularity, their adaptation is tied to active executions rather than a persistent worker composition \cite{wu2024loongserve, chen2026nanocp}.
Therefore, they do not determine how a fixed GPU budget should be distributed across CP degrees as aggregate demand evolves.
Moreover, transient rank groups hinder prefix reuse because they provide no stable workers for retaining cached prefixes and routing future requests.

To address these limitations, we propose \ourname{}\footnote{\ourname{} is the Roman god of change and transformation. 
It reflects the system's ability to adapt to evolving workloads across multiple timescales.}, an adaptive serving system that treats the CP worker composition as a cluster-wide, dynamically controllable resource.
\ourname{} coordinates CP execution at two timescales.
At the request level, its scheduler routes each incoming request among persistent CP workers by minimizing a model-based placement cost, which combines predicted queuing delay, cache-aware prefill time, and the GPU-time cost of the selected CP degree.
At the cluster level, \ourname{} monitors workload conditions and reconfigures the worker composition within seconds; 
it uses in-place split and merge operations that require neither process restarts nor model-weight reloads.
To maintain cache locality under heterogeneous and changing CP degrees, \ourname{} further provides a global prefix-cache manager that coordinates prefix placement, replication, and reclamation among workers with the same or different CP degrees.
Together, these mechanisms allow \ourname{} to improve serving capacity under a fixed GPU budget while satisfying service-level objectives (SLOs) under dynamic and prefix-intensive workloads.

In general, the paper makes the following contributions: 
\begin{itemize}[leftmargin=1em]
    \item We develop a cache-aware prefill-time model and incorporate the CP degree into request scheduling across heterogeneous workers.
    The scheduler jointly accounts for worker load, prefix reuse, CP-dependent performance, and GPU-time cost.

    \item We design a cluster-level controller that adjusts the CP worker composition as the workload changes.
    It splits and merges workers within seconds to better match the available parallel capacity to current serving demand.
    
    \item We design a global prefix-cache manager that preserves locality as request assignments and CP worker composition change.
    It adapts prefix placement and replication within and across CP degrees as reuse patterns evolve.

    \item We implement \ourname{} and evaluate it with public and production workloads.
    The results show that \ourname{} reduces mean and P90 TTFT by up to 28.1\% and 55.0\%, respectively, and improves token-weighted TTFT SLO attainment by up to 13.3 percentage points.
\end{itemize}

\section{Background and Motivation}
\label{sec:preliminary}

\subsection{Context Parallelism}

Modern LLMs employ diverse attention mechanisms, including dense, sparse, linear, and hybrid variants \cite{yang2025lserve, xu2026deepseek, team2026kimi, team2026qwen3}.
Context parallelism (CP) parallelizes long-sequence attention by partitioning the input tokens and their associated attention computation across multiple ranks, each mapped to one GPU.
Given an input sequence $\mathbf{x}=(x_0,\ldots,x_{n-1})$, standard causal self-attention \cite{vaswani2017attention} computes
\[
\operatorname{Attn}(Q,K,V) = \operatorname{softmax} \left(\frac{QK^\top}{\sqrt{d_h}} + M \right)V,
\]
where $Q$, $K$, and $V$ denote the query, key, and value representations, respectively; $d_h$ is the dimension of each attention head; and $M_{ij}=0$ for $j\leq i$ and $M_{ij}=-\infty$ otherwise.
The resulting valid attention region is triangular, with later query blocks involving more computation than earlier ones.

One implementation of CP addresses this imbalance through zigzag sequence partitioning \cite{jiang2025dcp, yang2025context}.
As illustrated in \autoref{fig:cp}, for a CP degree of $C$, the input sequence is padded, if necessary, to a multiple of $2C$ and evenly divided into $2C$ contiguous chunks.
Rank $k$ receives the $k$-th and $(2C-1-k)$-th chunks for $k=0,1,\ldots,C-1$, pairing an early chunk with a late chunk to balance the attention workload.
Each rank computes the query, key, and value representations for its local tokens and performs an \texttt{AllGather} to collect the key and value representations across all ranks.
It then computes attention for its local queries over all causally visible key-value states.
Finally, the local outputs are restored to their original sequence order and concatenated, producing a result equivalent to single-rank causal attention.
Besides this zigzag-partitioned design, several alternative CP execution schemes have also been proposed \cite{liu2024ringattention, brandon2023striped, fang2024usp, jacobs2023deepspeed}, and CP has been widely adopted in large-scale LLM training \cite{ge2025bytescale, gu2024loongtrain, li2023sequence, wang2025flexsp}.

\begin{figure*}
    \centering
    \subfigure[Long-request fraction]{
	\begin{minipage}{0.33\textwidth}{
			\includegraphics[width=1\textwidth]{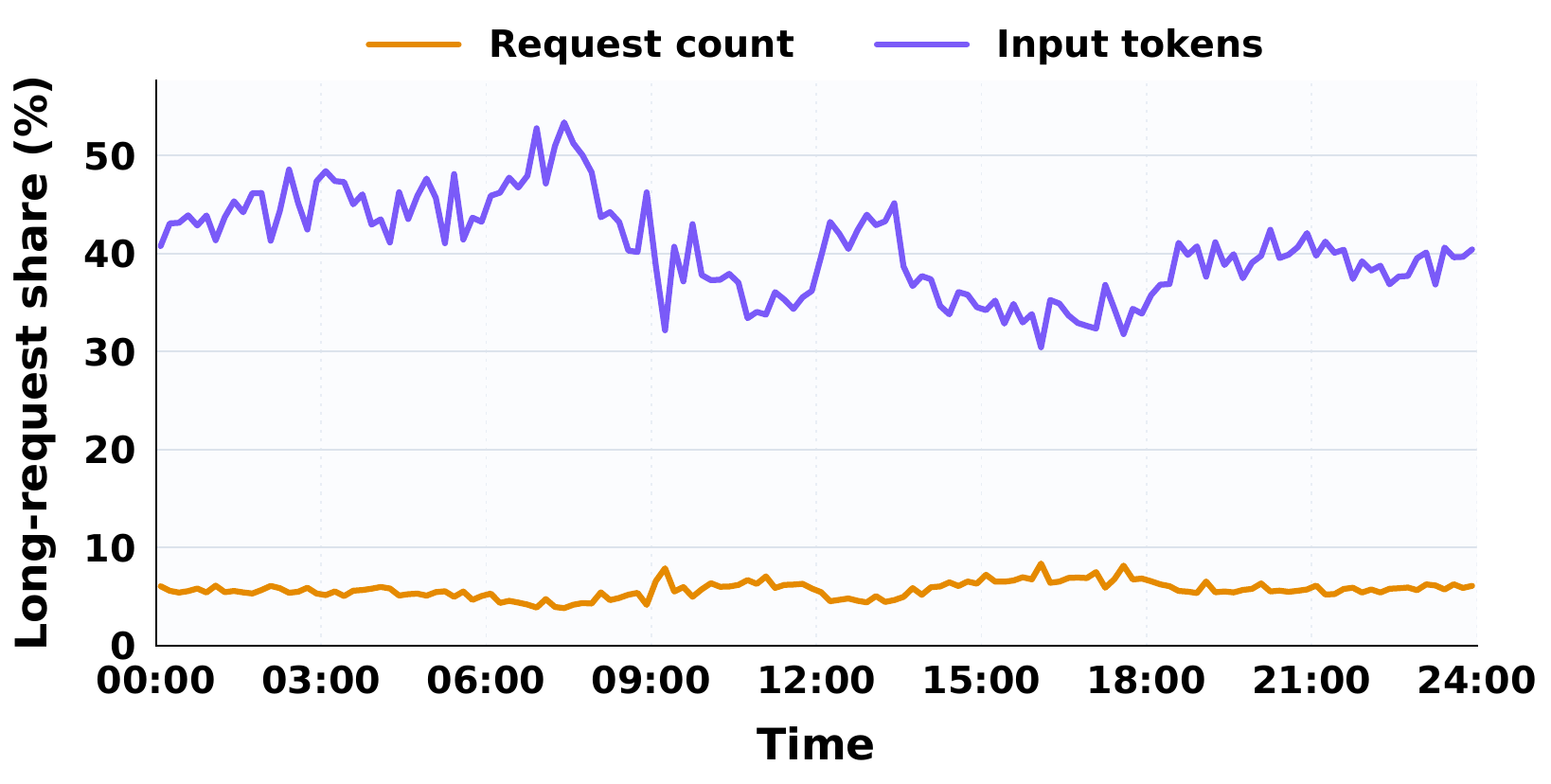}
            \label{fig:moti:short}
            \vspace{-0.1in}
        }
    \end{minipage}}%
    \subfigure[Normalized request rate]{
	\begin{minipage}{0.33\textwidth}{
			\includegraphics[width=1\textwidth]{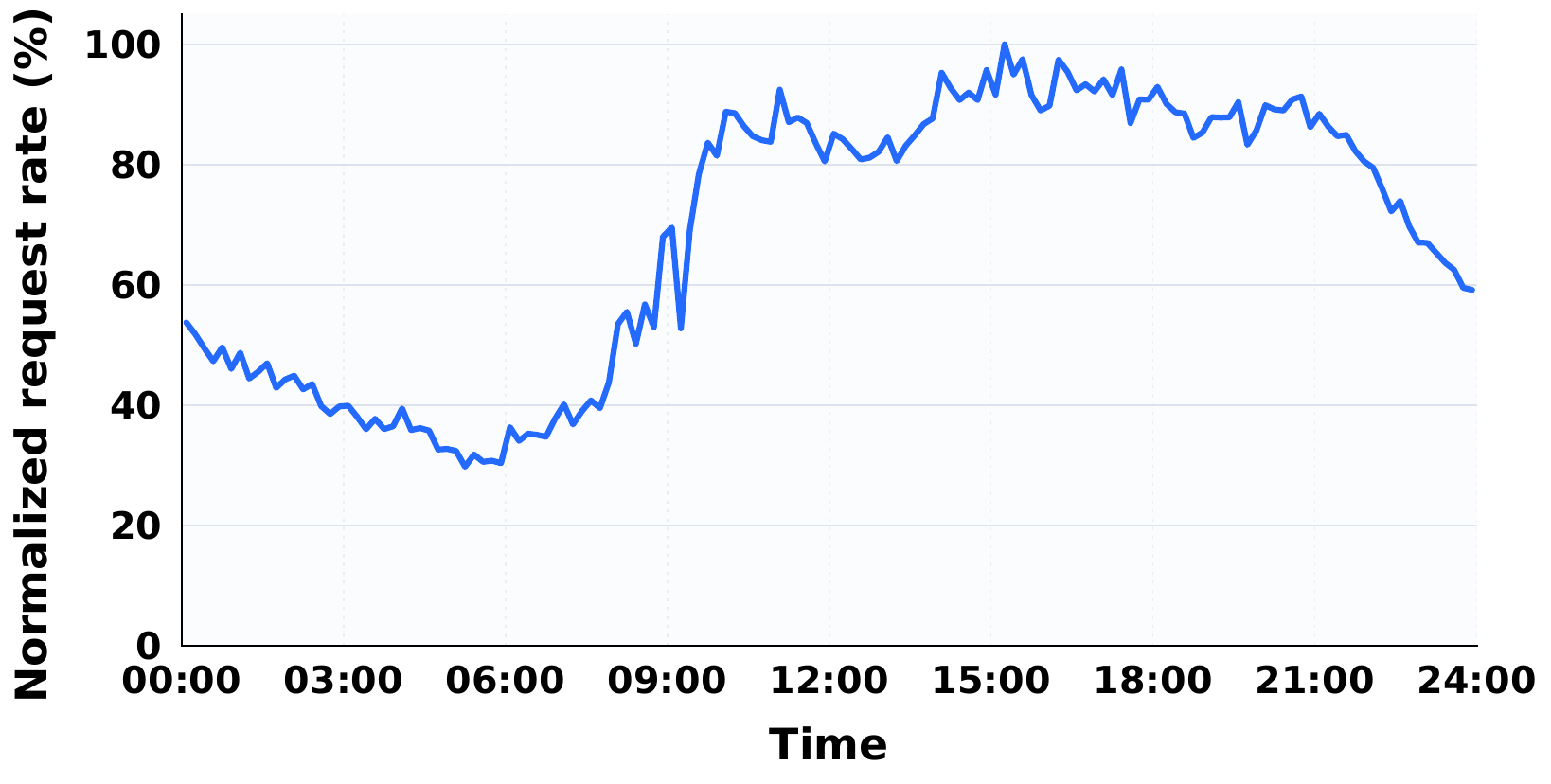}
            \label{fig:moti:qps}
            \vspace{-0.1in}
        }
    \end{minipage}}%
    \subfigure[Prefix-cache hit rate]{
	\begin{minipage}{0.33\textwidth}{
			\includegraphics[width=1\textwidth]{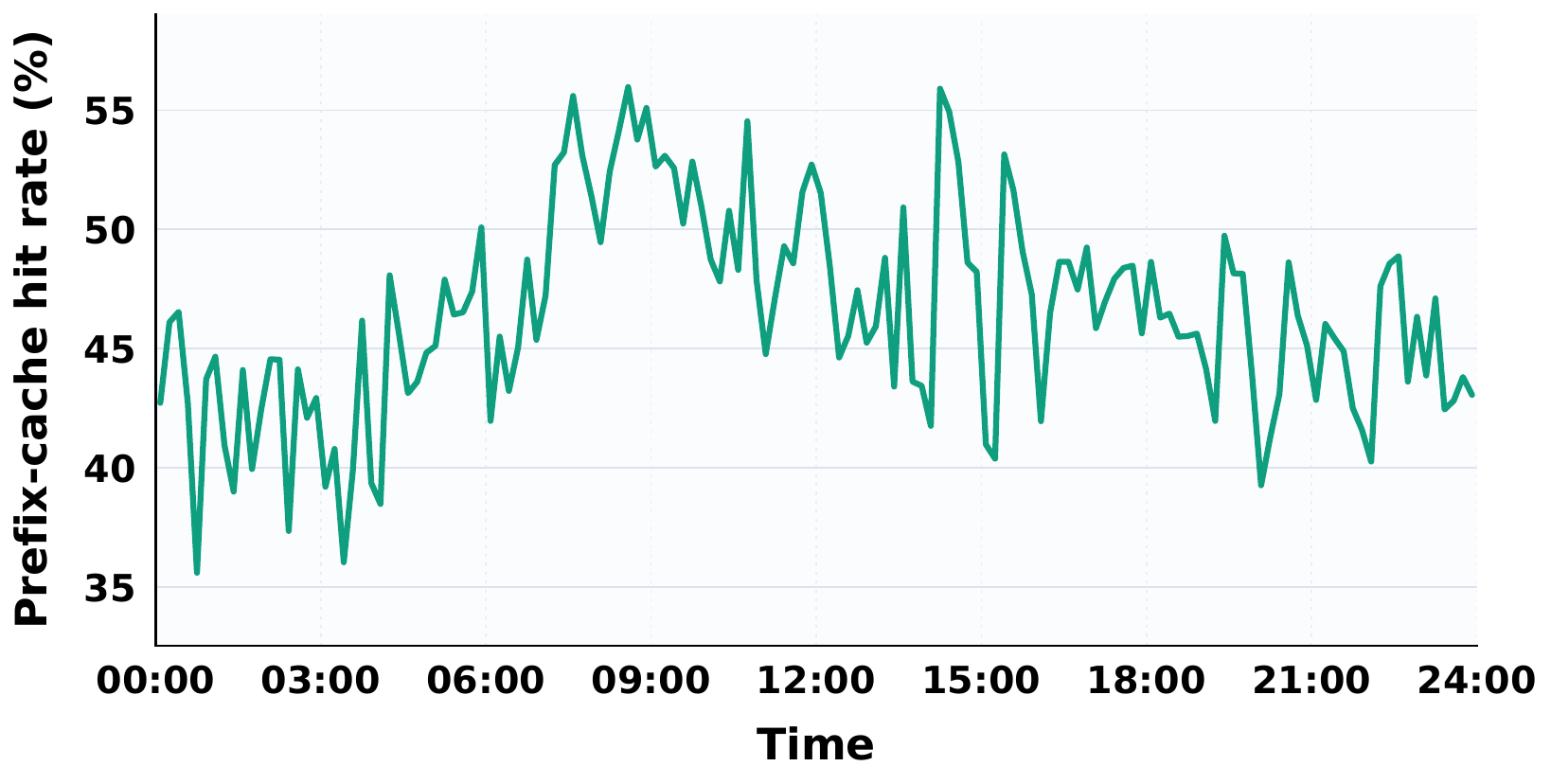}
            \label{fig:moti:cache}
            \vspace{-0.1in}
        }
    \end{minipage}} 
    \vspace{-0.1in}
    \caption{Temporal dynamics of a production LLM workload.}
    \label{fig:workload}
    \vspace{-0.15in}
\end{figure*}

For LLM inference, CP is particularly beneficial during the prefill stage, which processes the input sequence and constructs the KV cache before token generation.
By executing a request across the $C$ ranks, CP increases the aggregate compute resources available to the request, thereby reducing its prefill latency and potentially improving its TTFT.
However, a larger CP degree also introduces additional communication and occupies more GPUs.
Its benefit therefore depends strongly on input length: short inputs may not provide enough computation to amortize the communication overhead, whereas long inputs generally benefit more from the additional parallelism.
As shown in \autoref{fig:cp-effect}, we report the prefill speedup of different CP degrees relative to single-rank execution without CP (i.e., $C=1$) for the Qwen3-30B-A3B model \cite{yang2025qwen3}.
For short requests, CP provides little benefit and can even increase prefill time because its communication and synchronization overheads outweigh the reduction in attention computation.
Its benefit grows substantially with input length: at 64K tokens, $C=4$ and $C=8$ achieve approximately $2.71\times$ and $3.91\times$ speedups, respectively.
Thus, a larger CP degree can substantially reduce the prefill latency of long requests, but may consume additional GPUs without providing latency benefits for short requests, creating a fundamental trade-off between prefill latency and aggregate serving capacity.

\subsection{Prefix Caching under Context Parallelism}

LLM inference generally consists of two stages: prefill and decode.
The prefill stage processes the input sequence and generates the key-value (KV) states for its input tokens, while the decode stage autoregressively generates new tokens and appends their KV states to the existing cache \cite{zhong2024distserve, qin2024mooncake, zhang2025pqcache}.
KV caching retains these states so that each decode step does not need to recompute the representations of preceding tokens \cite{wang2025kvcache, shi2024keep}.
Prefix caching extends this reuse across requests, allowing requests in multi-turn interactions or requests sharing common prefixes to reuse previously computed KV states \cite{wang2026prefixkv, ye2024chunkattention}.
By retaining their KV states for future requests, prefix caching reduces redundant prefill computation at the cost of additional memory consumption \cite{li2025hotprefix, yang2026learned}.

Prefix caching reduces the computation for CP-based prefill.
Let $L=P+R$ denote the total input length, where $P$ is the cached prefix length and $R$ is the uncached suffix length.
Upon a prefix hit, the prefix KV states are reused, and only the $R$ uncached tokens undergo projection, feed-forward, and attention computation.
Because the uncached queries attend to both the cached prefix and preceding uncached tokens, the remaining attention workload is
\begin{equation}
    W(R, P)= RP+\frac{R(R+1)}{2}.
    \label{eq:remaining-attention}
\end{equation}
The two terms capture attention from the uncached queries to the cached prefix and causal attention within the uncached suffix, respectively.
As the cached prefix grows, CP has less remaining computation to parallelize, while its communication and synchronization overheads may not decrease proportionally.
Consequently, a larger CP degree may provide substantial benefits for an uncached request but only limited benefits for the same request after a large prefix hit.
Moreover, the cached KV states must be available at the selected worker and partitioned across its ranks, making their placement and reuse dependent on the worker's CP degree.
Therefore, selecting an appropriate CP degree requires considering both the cached prefix and the uncached suffix rather than the original input length alone.

\subsection{Motivating Observations}

To understand the workload characteristics encountered in deployment, we analyze a representative 24-hour traffic trace collected from our production LLM serving system.
\autoref{fig:workload} summarizes the temporal variation in its request-length composition, request arrival rate, and prefix-cache reuse.

\textbf{Observation 1: Long requests are sparse in count but large in token volume.}
We classify requests with more than 32K input tokens as long requests.
As shown in \autoref{fig:moti:short}, long requests account for approximately 4\% to 8\% of request arrivals, yet contribute about 32\% to 54\% of the input tokens over the day.
Long requests generally benefit more from larger CP degrees because they can better amortize CP communication overheads, whereas assigning larger degrees to short requests occupies more GPUs with limited latency benefit.
Consequently, request count alone understates the resource significance of long requests: even a small change in their frequency can materially change the aggregate prefill compute demand and the worker composition required to sustain it.

\textbf{Observation 2: Request arrival rate varies substantially over time.}
As shown in \autoref{fig:moti:qps}, the request arrival rate changes by more than threefold between off-peak and peak periods.
During off-peak periods, lower concurrency pressure allows more GPUs to be assigned to individual requests to reduce their latency.
During peak periods, however, a worker composition containing too many larger-degree workers provides fewer independent serving lanes and may limit aggregate serving capacity.
Consequently, a worker composition that is effective under light load may become a bottleneck as the request arrival rate increases.

\textbf{Observation 3: Prefix-cache reuse is common and dynamic.}
We define the token-level prefix-cache hit rate as the fraction of input tokens covered by prefix-cache hits.
As shown in \autoref{fig:moti:cache}, this hit rate varies between approximately 35\% and 55\% over the day.
Prefix hits reduce the amount of computation remaining for CP and change the relative benefit of different CP degrees.
Moreover, cache contents are worker-specific, even among workers with the same CP degree, while changes in worker composition may require cached KV states to be transferred or reconciled across workers.
Request placement and worker reconfiguration must therefore both account for the current prefix-cache state.

\section{Problem Definition}
\label{sec:prob_def}

\subsection{System Model}

We consider the prefill pool of a P/D-disaggregated LLM serving system with a fixed budget of $G$ GPUs.
A persistent serving unit with CP degree $C$ is referred to as a CP worker, or simply a worker.
A degree-$C$ worker consists of $C$ CP ranks, each mapped to one GPU, and serves as the unit of request execution and physical KV-state placement.
At time $t$, the prefill pool contains a set of workers $\mathcal W(t)$.
Each worker $w$ has a CP degree $C_w\in\mathcal C$ and occupies $C_w$ GPUs:
\[
\sum_{w\in\mathcal W(t)} C_w = G,
\]
where $\mathcal C$ is the predefined set of supported CP degrees.
Each worker's CP degree remains fixed over the request-scheduling timescale.

Consider a request $i$ arriving at time $t_i$ with input length $L_i$.
Let $P_{i,w}(t_i)$ denote the longest reusable prefix available at worker $w$, and let
\[
R_{i,w}(t_i)=L_i-P_{i,w}(t_i)
\]
be the remaining uncached suffix.
We denote the prefill execution time under CP degree $C$ by $E_C(R,P)$.
If request $i$ is assigned to worker $w$, its time-to-first-token (TTFT) is
\[
\operatorname{TTFT}_{i,w}
=
Q_w(t_i)
+
E_{C_w}\!\left(R_{i,w}(t_i),P_{i,w}(t_i)\right)
+
T_i^{\mathrm{ext}},
\]
where $Q_w(t_i)$ is its waiting time at worker $w$, and $T_i^{\mathrm{ext}}$ captures latency outside the prefill pool, including KV transfer and first-token generation.
Let $w_i$ denote the selected worker, such that $\operatorname{TTFT}_i=\operatorname{TTFT}_{i,w_i}$.
After prefill, requests and their KV states are transferred to a separately provisioned decode pool for token generation.

\subsection{Cache-Aware Prefill-Time Model}

Building on profiling-based models for sequence-parallel prefill and cache-aware routing
\cite{wu2024loongserve, srivatsa2025preble}, we construct a compact model
that jointly captures CP degree and prefix reuse.
With $P$ denoting the cached prefix length and $R$ the uncached suffix length, the predicted execution time is modeled as
\begin{equation}
\label{eq:prefill-performance-model}
\widehat E_C(R,P)
=
d+aR+bP
+
\frac{1}{C}
\left[
cR+\alpha W(R,P)
\right].
\end{equation}
Here, $d$ is the fixed startup overhead, and $aR$ captures uncached-token costs that do not decrease proportionally with CP, including communication, synchronization, and data movement.
The term $bP$ captures cached-prefix access and replay, while $cR$ represents token-local computation distributed across CP ranks, such as projection and feed-forward operations.
Finally, $W(R,P)=RP+R(R+1)/2$ is the remaining attention workload derived in \autoref{eq:remaining-attention}, and $\alpha$ converts it into execution time.
The coefficients are fitted from queue-free profiles for each model-hardware configuration.

\begin{figure}
    \centering
    \subfigure[Predicted vs. measured prefill time]{
	\begin{minipage}{0.22\textwidth}{
			\includegraphics[width=1\textwidth]{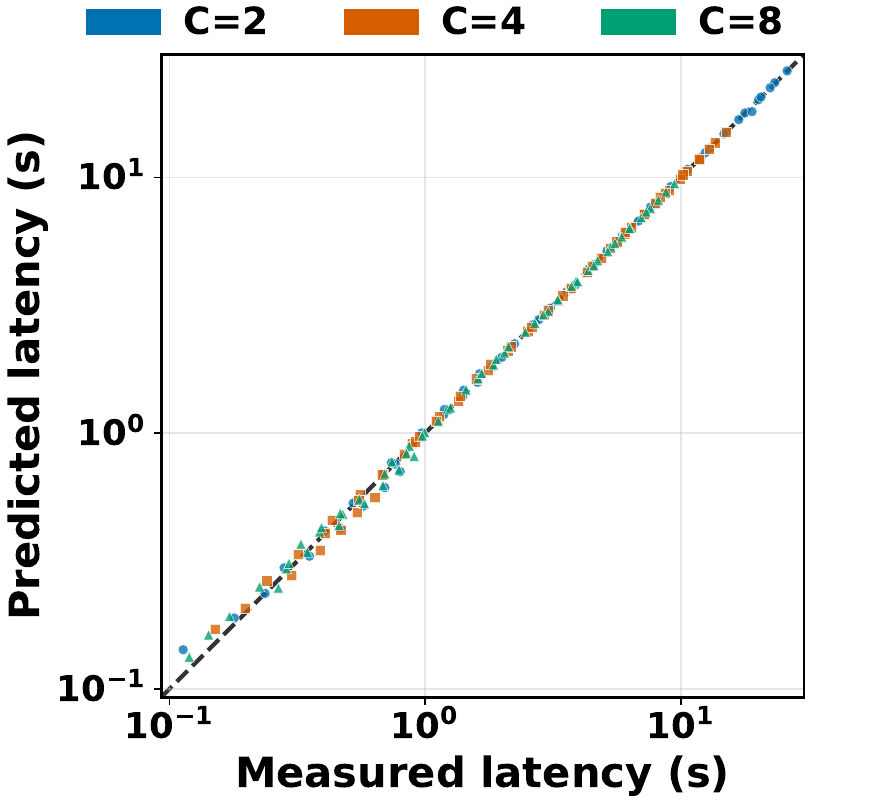}
            \vspace{-0.1in}
        }
    \end{minipage}}%
    \subfigure[Prediction error distribution]{
	\begin{minipage}{0.22\textwidth}{
			\includegraphics[width=1\textwidth]{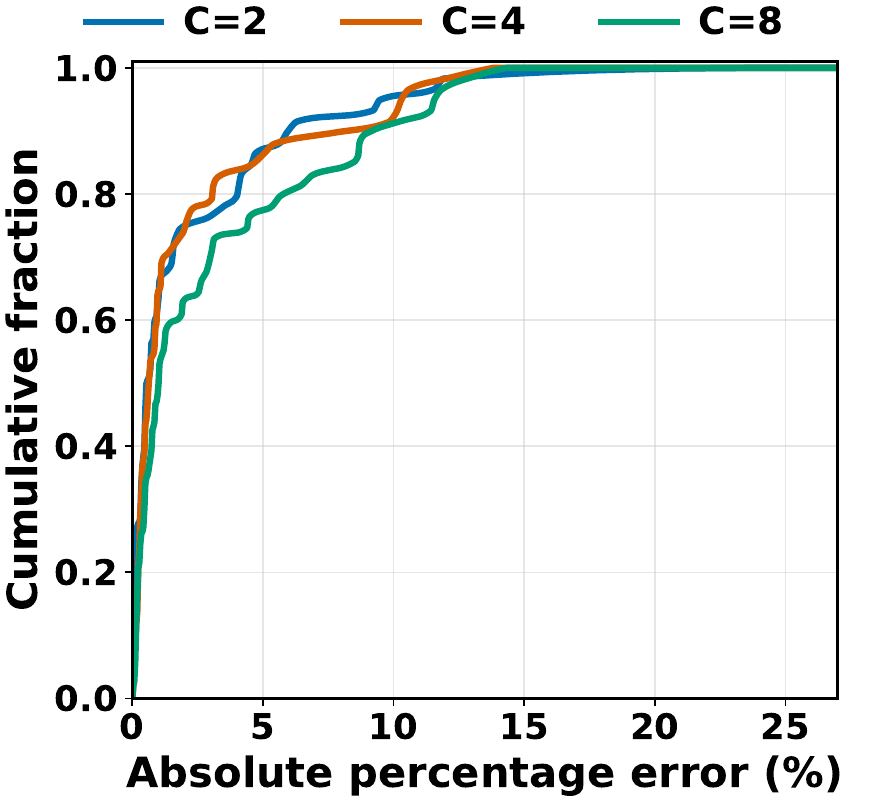}
            \vspace{-0.1in}
        }
    \end{minipage}}
    \vspace{-0.1in}
    \caption{Accuracy of the cache-aware prefill-time model.}
    \label{fig:prefill-model-error}
    \vspace{-0.2in}
\end{figure}

This model captures the joint effects of prefix reuse and context parallelism.
For a fixed input length, a larger prefix hit reduces both $R$ and $W(R,P)$, while increasing $C$ accelerates only the computation distributed across CP ranks.
The benefit of a CP degree therefore depends on the request's remaining computation rather than its original input length alone.
As shown in \autoref{fig:prefill-model-error}, the model yields small fitting errors for the Qwen3-30B-A3B model \cite{yang2025qwen3} across the profiled CP degrees, input lengths, and prefix-cache hit ratios, indicating that this compact decomposition captures the dominant prefill costs.

For a request-worker pair $(i,w)$, we instantiate the model as
\begin{equation}
    s_{i,w}= \widehat E_{C_w}
\!\left(R_{i,w}(t_i),P_{i,w}(t_i)\right),
\label{eq:prefill-estimate}
\end{equation}
which denotes the predicted prefill service time excluding queuing delay.

\subsection{Service Objectives}

For each request $i$, we specify a TTFT requirement $\tau_i$ and a time-per-output-token (TPOT) requirement $\delta_i$.
For a measurement window of duration $H$, let $\mathcal R(H)$ denote the requests arriving during the window; their latencies are measured upon completion.
Since \ourname{} operates on the prefill pool, its primary objective is to minimize mean TTFT:
\[
\operatorname{TTFT}(H)
=
\frac{1}{|\mathcal R(H)|}
\sum_{i\in\mathcal R(H)}
\operatorname{TTFT}_i.
\]
It also seeks to maximize token-weighted TTFT SLO attainment:
\[
\operatorname{Attain}_{\mathrm{TTFT}}(H)=\frac{
\sum_{i\in\mathcal R(H)}L_i \cdot 
\mathbf 1\left[\operatorname{TTFT}_i\leq\tau_i\right]}{
\sum_{i\in\mathcal R(H)}L_i }.
\]
Here, each request is weighted by its input length $L_i$, so the metric measures the fraction of input-token demand contributed by requests that satisfy their TTFT requirements.

We assume that the decode pool is independently managed to meet its TPOT requirements.
Improving prefill efficiency may allow the system to sustain a higher request rate and consequently increase decode throughput when sufficient decode capacity is available, while decode-side optimization is outside the scope of this work.

\section{System Overview}
\label{sec:overview}

\begin{figure}
    \centering
    \includegraphics[width=.9\linewidth]{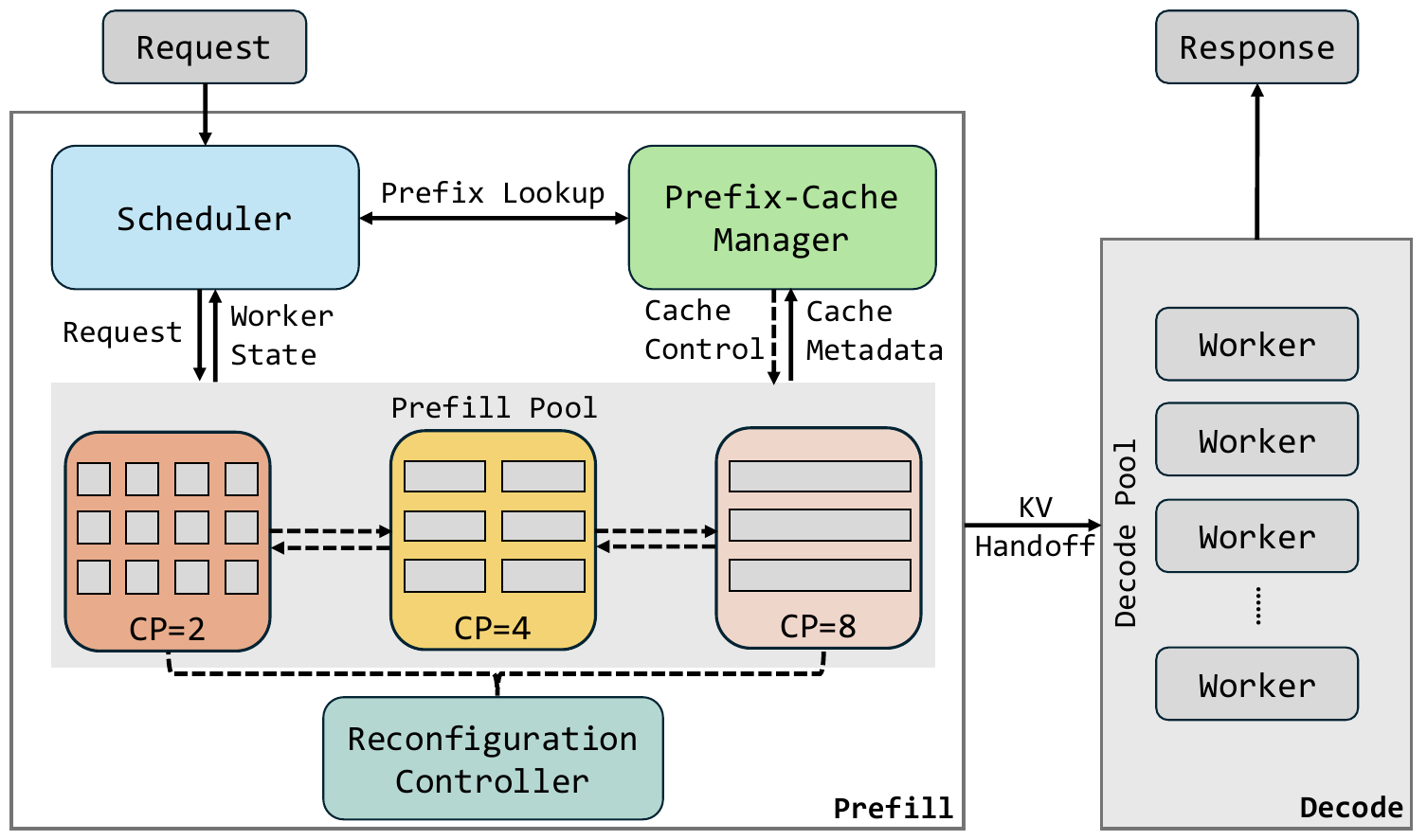}
    \vspace{-0.1in}
    \caption{System architecture of \ourname{}. Solid arrows denote request-level data and control paths, whereas dashed arrows represent cluster-level management operations.}
    \label{fig:overview}
    \vspace{-0.1in}
\end{figure}

To address these challenges, we present \ourname{}, an LLM serving system that adapts context parallelism to heterogeneous and dynamically changing production workloads.
\autoref{fig:overview} illustrates its overall architecture.
\ourname{} operates within the prefill pool of a P/D-disaggregated serving system: a fixed GPU budget is partitioned among persistent CP workers with different CP degrees, while requests are subsequently transferred to an independently provisioned decode pool.
Within the prefill pool, \ourname{} coordinates request placement, worker composition, and prefix-cache management to reduce mean TTFT and improve TTFT SLO attainment under a fixed GPU budget.

\ourname{} adapts CP at two complementary timescales.
At the request timescale, the system decides which existing CP worker should serve each request; at the cluster timescale, it decides which CP workers should exist.
For each incoming request, the scheduler consults the current worker state and worker-specific prefix-cache information to select an appropriate worker from the eligible workers with different CP degrees.
Over longer workload windows, the reconfiguration controller monitors workload conditions and adjusts the worker composition through worker split and merge operations.
Across both timescales, the global prefix-cache manager adapts prefix placement to request routing and worker reconfiguration.

\section{Request-Level Scheduling}
\label{sec:schedule}

Given the current worker composition and prefix-cache placement, the request-level scheduler assigns each incoming request to an eligible CP worker.
This decision jointly considers request length, worker load, and worker-specific prefix reuse across different CP degrees.
In this section, we first explain why considering any one of these factors alone is insufficient, then describe the worker state and candidate construction, and finally present the CP-aware placement cost used by \ourname{}.

\subsection{Scheduling Challenges}

Under a fixed worker composition, workers with different CP degrees are not interchangeable.
Assigning a request to a worker with a larger CP degree may reduce its prefill latency, but it also occupies more GPUs and limits the capacity available for serving other requests concurrently.
This resource opportunity cost means that minimizing individual request latency does not necessarily maximize aggregate input-token throughput.
Prefix reuse further changes the computation remaining for a request and, consequently, the CP degree at which it can be served efficiently.
Requests with the same original input length may therefore favor different CP degrees depending on their available prefix reuse, making length-only routing insufficient.

These considerations must also be balanced against the current worker load.
A cache-first policy may repeatedly select an overloaded worker, increasing queuing delay despite avoiding redundant computation.
Conversely, a load-first policy may select an idle worker but require recomputation of a large reusable prefix.
Routing based only on request length and CP degree ignores both load imbalance and worker-specific prefix locality.
A CP-aware scheduler must therefore jointly consider request length, worker load, and prefix-cache reuse.

\subsection{Worker State and Candidate Construction}

When request $i$ arrives at time $t_i$, the scheduler observes the current worker set $\mathcal W(t_i)$ and three types of state for each worker $w\in\mathcal W(t_i)$: its CP degree $C_w$, its outstanding prefill workload, and its lifecycle status.
The scheduler uses this workload as a lightweight proxy for the waiting time $Q_w(t_i)$, which depends on subsequent execution progress and is not directly available at placement time.
This proxy enables the scheduler to compare load pressure across workers with different CP degrees.
The scheduler also queries the prefix-cache manager for the longest reusable prefix of request $i$ available at each worker.
Because cache contents are worker-specific, these lengths may differ across candidate workers for the same request.

Using this state, the scheduler constructs the candidate set
\[
\mathcal W_i(t_i)
=
\left\{
w\in\mathcal W(t_i)
\mid
w\text{ is eligible to serve request }i
\right\}.
\]
A worker is excluded if it is unavailable or draining, or if it cannot satisfy the sequence-length or memory requirements of the request.
A prefix-cache miss does not exclude a worker, since the worker can still serve the request by recomputing the missing prefix.
The resulting candidate set and its worker-specific state are used to evaluate the placement costs described in the next subsection.
Once a worker is selected, the scheduler immediately reserves the request's estimated prefill workload at that worker so that concurrent arrivals do not observe the same pre-assignment load state.

\subsection{Adaptive Placement Cost}

To select a worker for request $i$, the scheduler assigns a placement cost to every eligible worker $w\in\mathcal W_i(t_i)$:
\[
J_{i,w}
=
q_w(t_i)
+
\beta s_{i,w}
+
\lambda\left[C_w s_{i,w} -C_{\min} \widehat E_{C_{\min}}(R_{i,w}(t_i), P_{i,w}(t_i))\right],
\]
where $C_{\min}$ is the smallest CP degree among the eligible workers, $\beta\geq1$ controls the emphasis on cache-aware service time, and $\lambda\geq 0$ controls the cost of occupying additional GPUs.
The request is assigned to the minimum-cost worker:
\[
w_i
=
\underset{w\in\mathcal W_i(t_i)}{\arg\min}\;
J_{i,w}.
\]

\noindent\textbf{Queuing-delay cost.}
The first term, $q_w(t_i)$, estimates how long request $i$ would wait before beginning execution at worker $w$.
It is maintained from the predicted remaining service times of the unfinished requests already assigned to worker $w$ at time $t_i$.
We estimate this delay as
\[
q_w(t_i)
=
\sum_{\substack{j:\,w_j=w\\j\text{ is unfinished at }t_i}}
\widehat E_{C_w}
\left(
R_{j,w}(t_j),
P_{j,w}(t_j)
\right).
\]
This term avoids routing requests to overloaded workers solely because they provide a favorable CP degree or prefix-cache hit.

\noindent\textbf{Service-time cost.}
The second term uses the performance model defined in \autoref{eq:prefill-estimate}:
\[
s_{i,w}
=
\widehat E_{C_w}
\left(R_{i,w}(t_i),P_{i,w}(t_i)\right).
\]
It captures the remaining computation, reusable prefix, and execution efficiency under the worker's CP degree.
When $\beta=1$, $q_w(t_i)+s_{i,w}$ corresponds to the predicted worker-dependent component of TTFT.
However, latency-greedy placement may select a lightly loaded worker without the reusable prefix, causing redundant computation and dispersing related prefixes across workers.
Setting $\beta>1$ amplifies the smaller service time produced by a prefix hit, making cache-holding workers more likely to be selected and preserving prefix locality for subsequent requests.

\noindent\textbf{Additional GPU-time cost.}
The final term measures the additional GPU time incurred by executing request $i$ at worker $w$ rather than at the smallest eligible CP degree.
Execution at worker $w$ is predicted to consume $C_ws_{i,w}$ GPU-seconds, whereas execution at degree $C_{\min}$ under the same prefix-reuse condition would consume
\[
C_{\min} \cdot \widehat E_{C_{\min}}
\left(R_{i,w}(t_i),P_{i,w}(t_i)\right)
\]
GPU-seconds.
Their difference accounts for both the larger GPU footprint and the shorter execution time of degree $C_w$, thereby capturing the resource cost of using additional CP ranks.
The coefficient $\lambda$ controls how strongly this cost affects request placement.
For short requests, limited parallel speedup typically results in greater excess GPU time and favors a smaller CP degree.
For compute-intensive requests, the larger latency reduction can outweigh this penalty, allowing them to select a larger degree.

\section{Cluster-Level Reconfiguration}

Request-level scheduling can route requests only among the currently available CP workers:
it cannot correct a persistent mismatch between the worker composition and aggregate workload demand.
To address this limitation, \ourname{} initializes a heterogeneous worker composition from historical workload characteristics and periodically adapts it through adjacent worker split and merge operations.
These transitions activate pre-established CP groups over persistent GPU ranks, avoiding process restarts and model-weight reloading.

\subsection{Capacity Profiling and Initialization}

Both offline profiling and runtime monitoring partition their measurements into windows of fixed duration $H$.
Before deployment, \ourname{} profiles one worker at each supported CP degree under increasing request loads.
For a degree-$C$ worker, the window-level request demand is measured by summing the predicted service times $\widehat E_C(R,P)$ of the requests served within an observation window.
We denote by $\kappa_C$ the largest window-level demand for which the worker still satisfies the target TTFT SLO attainment.
This profiling is performed for each model and CP degree deployed in our testbed and captures their execution efficiency and runtime overheads.

The initial worker composition is selected using a representative historical workload, including its aggregate demand and request-length distribution.
Historical information determines the expected mixture of small- and large-degree workers, while the profiled capacities are used to ensure that the selected composition can accommodate the expected demand.
Among historically suitable compositions with sufficient capacity, the controller favors compositions containing more large-degree workers, making greater parallelism available to latency-sensitive, compute-intensive requests.
If none of these compositions provides sufficient capacity, it selects the composition with the largest aggregate profiled capacity.

The runtime windows are consecutive and non-overlapping, with $H_k$ denoting the $k$-th window.
Let $\mathcal R(H_k)$ contain the requests arriving during $H_k$, and let $w_i$ denote the worker selected for request $i$ in that window.
The aggregate workload demand is estimated as
\[
D^{(k)}
=\sum_{i\in\mathcal R(H_k)}
\widehat E_{C_{w_i}}\left(R_{i,w_i}(t_i),P_{i,w_i}(t_i)\right).
\]
The controller uses this estimate to determine whether the current composition should remain unchanged or move to an adjacent state in the pre-established grouping lattice.

\subsection{Split-and-Merge Reconfiguration}

At the end of each observation window, the controller compares the aggregate workload demand with the profiled capacity of the current worker composition and also monitors the fraction of long requests.
A sustained workload above the upper threshold triggers a split to increase serving concurrency, whereas a workload below the lower threshold triggers a merge to provide more parallelism for individual requests.
When the aggregate workload remains relatively stable but the long-request fraction changes substantially, an increase in this fraction favors merging, while a decrease favors splitting.
Capacity takes precedence when the two signals conflict, and a merge is admitted only if the adjacent composition retains sufficient profiled capacity for the observed workload.
The corresponding condition must persist for multiple observation windows before a transition is initiated.

Given the selected direction, the controller considers topology-compatible split or merge operations leading to an adjacent composition.
Each operation replaces one degree-$2C$ worker with two degree-$C$ workers or performs the inverse conversion, thereby preserving the aggregate GPU count.
Among eligible operations, the controller first minimizes the predicted drain interval and then the expected KV-reconciliation traffic.
The latter is estimated from the missing destination copies of retained prefixes weighted by their recent access frequencies.

The participating workers are removed from the scheduler's candidate set and stop admitting new prefill requests.
Queued requests are redirected to other eligible workers, while in-flight prefill executions finish under the old CP geometry.
The CP geometry is pinned throughout each engine step, and a configuration update observed at the model-input boundary applies only to the next step.
This prevents a single execution step from combining communication under the old geometry with sequence partitioning under the new geometry.

After a parent worker has drained, a split can activate its independent child groups.
For a merge, all participating sibling groups first park at a common model-input boundary before activating their parent group.
The controller commits the target CP degree under a new configuration epoch, and the resulting workers become schedulable only after all participating ranks have adopted the new epoch and the scheduler and cache manager have updated their worker metadata.

Cached prefixes are retained without placing their full redistribution on the conversion critical path.
Although their physical KV blocks remain resident, their rank ownership and addressing may need to be reconciled with the new CP geometry.
After a merge, the manager combines the prefix metadata of the source workers and marks blocks missing from the merged worker as pending reconciliation.
After a split, it assigns retained prefixes to the resulting workers according to prefix demand and available cache capacity.
When a pending prefix is first reused, the manager transfers only the missing blocks from a valid resident copy and installs them under the new CP geometry.
A pending prefix is not counted as immediately reusable cache capacity until reconciliation completes, and its missing portion is processed as an ordinary cache miss if no valid copy remains.
Subsequent replication and reclamation are handled by the prefix-cache manager described in the next section.

\section{Prefix Cache Management}

\begin{figure}
    \centering
    \includegraphics[width=\linewidth]{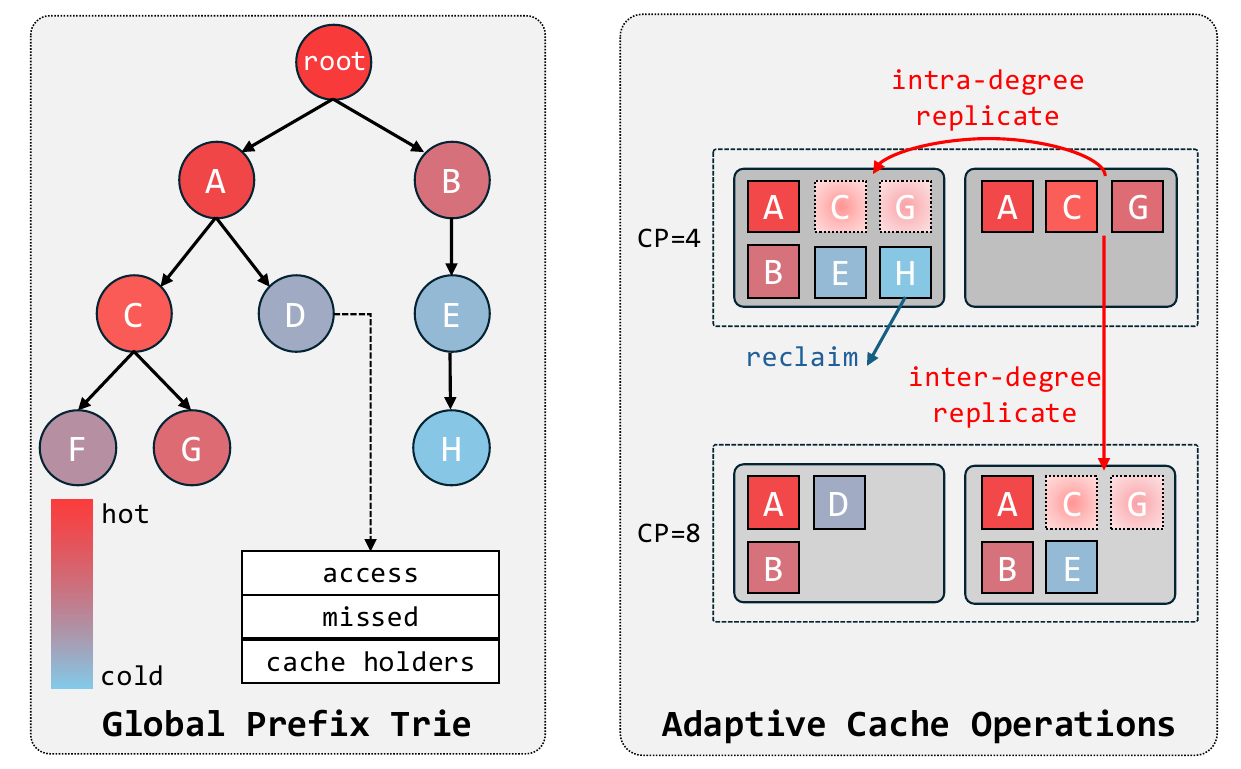}
    \vspace{-0.25in}
    \caption{Prefix-cache management in \ourname{}. Colors indicate recent demand. Each gray container represents the local prefix cache of a CP worker, while blocks with dotted outlines denote replicas created after a periodic scan.}
    \label{fig:cache}
    \vspace{-0.15in}
\end{figure}

\ourname{} introduces a global prefix-cache manager to preserve cache locality across heterogeneous CP workers and worker reconfiguration.
It maintains a global view of the logical prefix structure, replica placement, and degree-level prefix demand.
Based on this information, the manager replicates cached prefixes within and across CP degrees and reclaims underused replicas.
\autoref{fig:cache} illustrates the global prefix trie and the resulting cache operations.

\subsection{Global Prefix-Trie Management}

Each CP worker maintains a physically separate prefix cache, and its KV states cannot be reused by another worker without explicit replication or transfer.
Request placement and worker reconfiguration continuously change where cached prefixes are accessed and stored, making it difficult for independent worker caches to preserve prefix locality.
Thus, \ourname{} maintains a global metadata view over the caches of all CP workers.
Following prior prefix-caching systems \cite{zheng2024sglang, ye2024chunkattention, srivatsa2025preble, hu2024memserve}, the manager organizes this metadata as a prefix trie to capture the inclusion relationships among cached prefixes.

In the prefix trie, each node $z$ represents one cache block, while the path from the root to $z$ represents the complete prefix ending at that block.
The manager also records its prefix length and the workers that currently hold the complete cached prefix ending at node $z$.
From these replica holders and their CP degrees, the manager derives $\mathtt{ReplicaCount}(z,C)$, which is the number of degree-$C$ workers holding this prefix.
For every supported CP degree $C$, each node additionally maintains two dynamic statistics, $\mathtt{access}[z,C]$ and $\mathtt{missed}[z,C]$.
$\mathtt{access}[z,C]$ records how often node $z$ is reused by requests assigned to degree-$C$ workers.
$\mathtt{missed}[z,C]$ accumulates the cache benefit lost when requests prefer degree $C$ but no degree-$C$ worker holds $z$.
The former captures demand for adjusting the number of existing replicas within a CP degree, whereas the latter reveals demand for introducing a replica at a CP degree where the prefix is currently unavailable.
The trie maintains these logical relationships and global metadata, while each worker's local cache index continues to track its physical KV blocks.

When request $i$ arrives, the manager traverses the trie using its block sequence to obtain $\mathtt{MatchedPath}(i)$, regardless of where the matched nodes are replicated.
After the scheduler assigns the request to worker $w_i$, $\mathtt{ReusedPath}(i,w_i)$ identifies the portion of the matched path whose KV states are actually reused at that worker.
The manager increments $\mathtt{access}[z,C_{w_i}]$ for every node on the reused path.
Because the selected worker may be affected by transient queuing pressure, the manager derives the request's preferred CP degree by omitting the worker-load term $q_w(t_i)$ from the placement cost:
\[
w_i^{\mathrm{pref}}
=
\underset{w\in\mathcal W_i(t_i)}{\arg\min}
\left[
J_{i,w}-q_{w}(t_i)
\right],
\qquad
C_i^{\mathrm{pref}}=C_{w_i^{\mathrm{pref}}}.
\]
We denote this degree by $\mathtt{PreferDegree}(i)=C_i^{\mathrm{pref}}$.
For each node on the matched path with no replica at $C_i^{\mathrm{pref}}$, the manager computes its marginal cache benefit as the difference between the predicted service times when prefix reuse ends at its parent and at the node itself.
This benefit is accumulated in $\mathtt{missed}[z,C_i^{\mathrm{pref}}]$.
Because each node contributes only the benefit of extending its parent prefix by one cache block, the benefits of longer prefixes are not counted repeatedly.
\autoref{alg:cache-update} summarizes this per-request update and gives the corresponding calculation.

\begin{algorithm}[!ht]
\caption{Per-Request Cache Statistics Update}
\label{alg:cache-update}
\begin{algorithmic}[1]
\Require Request $i$ and its selected worker $w_i$
\ForAll{$z\in\mathtt{ReusedPath}(i,w_i)$}
    \State Increase $\mathtt{access}[z,C_{w_i}]$ by $1$
\EndFor

\State $C_i^{\mathrm{pref}} \gets \mathtt{PreferDegree}(i)$

\ForAll{$z\in\mathtt{MatchedPath}(i)$}
    \If{$\mathtt{ReplicaCount}(z,C_i^{\mathrm{pref}})=0$}
        \State $P_0\gets
        \mathtt{PrefixLength}(\mathtt{Parent}(z))$
        \State $P_1\gets\mathtt{PrefixLength}(z)$
        \State $\mathtt{benefit}\gets
        \left[\widehat E_{C_i^{\mathrm{pref}}}(L_i-P_0,P_0)-\widehat E_{C_i^{\mathrm{pref}}}(L_i-P_1,P_1)
        \right]$
        \State Increase
        $\mathtt{missed}[z,C_i^{\mathrm{pref}}]$
        by $\mathtt{benefit}$
    \EndIf
\EndFor
\end{algorithmic}
\end{algorithm}

\subsection{Adaptive Replication and Reclamation}

The cache manager periodically scans every trie node $z$ at each supported CP degree $C$ and generates inter-degree replication, intra-degree replication, and reclamation candidates.
Inter-degree replication creates the first replica of $z$ at degree $C$ when none currently exists.
When $\mathtt{ReplicaCount}(z,C)=0$, node $z$ cannot produce cache accesses at degree $C$, so $\mathtt{access}[z,C]$ cannot capture the demand for introducing such a replica.
The manager therefore uses $\mathtt{missed}[z,C]$, which accumulates the predicted cache benefit lost by requests that prefer degree $C$, and adds $(z,C)$ to the inter-degree replication list when this value exceeds a threshold $\theta_{\mathrm{inter}}$.

When $r=\mathtt{ReplicaCount}(z,C)>0$, intra-degree replication creates an additional replica among workers with the same CP degree.
In this case, $\mathtt{access}[z,C]/r$ represents the average reuse demand handled by each existing replica.
The manager adds $(z,C)$ to the intra-degree replication list when $\mathtt{access}[z,C]>r\theta_{\mathrm{intra}}$, where $\theta_{\mathrm{intra}}$ represents the per-replica demand threshold required to justify another same-degree copy.
Conversely, it adds $(z,C)$ to the reclamation list when $\mathtt{access}[z,C]<r\theta_{\mathrm{reclaim}}$, indicating that the existing replicas no longer justify their cache occupancy.
We set $\theta_{\mathrm{reclaim}}<\theta_{\mathrm{intra}}$ to avoid repeated replication and reclamation.
The thresholds $\theta_{\mathrm{inter}}$, $\theta_{\mathrm{intra}}$, and $\theta_{\mathrm{reclaim}}$ are calibrated before deployment for the target model, hardware configuration, and scan interval.
After each scan, the manager multiplies both $\mathtt{access}$ and $\mathtt{missed}$ by a configured factor $\mathtt{decay}\in(0,1)$, giving recent demand greater influence than older observations.
\autoref{alg:cache-scan} summarizes the periodic scan.

\begin{algorithm}[!ht]
\caption{Periodic Prefix-Cache Scan}
\label{alg:cache-scan}
\begin{algorithmic}[1]
\Require Prefix trie, supported CP degrees $\mathcal C$

\State Initialize empty lists $\mathtt{inter}$, $\mathtt{intra}$, and $\mathtt{reclaim}$

\ForAll{node $z$ and degree $C\in\mathcal C$}
    \State $r\gets\mathtt{ReplicaCount}(z,C)$

    \If{$r=0$}
        \If{$\mathtt{missed}[z,C]>
            \theta_{\mathrm{inter}}$}
            \State Add $(z,C)$ to $\mathtt{inter}$
        \EndIf
    \Else
        \If{$\mathtt{access}[z,C]>r\theta_{\mathrm{intra}}$}
            \State Add $(z,C)$ to $\mathtt{intra}$
        \ElsIf{$\mathtt{access}[z,C]< r\theta_{\mathrm{reclaim}}$}
            \State Add $(z,C)$ to $\mathtt{reclaim}$
        \EndIf
    \EndIf
\EndFor

\ForAll{node $z$ and degree $C\in\mathcal C$}
    \State Scale $\mathtt{access}[z,C]$ by $\mathtt{decay}$
    \State Scale $\mathtt{missed}[z,C]$ by $\mathtt{decay}$
\EndFor
\end{algorithmic}
\end{algorithm}

The manager first determines which candidates in the three lists should be executed.
For each replication candidate, the manager estimates the minimum copy footprint among feasible destination workers.
For a given destination, this footprint contains only the missing blocks on the path from the root to $z$.
Inter-degree candidates are ranked by decreasing $\mathtt{missed}[z,C]$ per copied block.
Intra-degree candidates are ranked by decreasing $\mathtt{access}[z,C]/r$, with smaller copy footprints preferred when their per-replica access frequencies are similar.
Reclamation candidates are ranked by increasing $\mathtt{access}[z,C]/r$.
The three lists are ranked separately, and only a bounded number of operations are selected from each list per scan.

For each selected replication, the destination is chosen from degree-$C$ workers without $z$.
The manager favors destinations that retain a longer ancestor path of $z$, have sufficient cache capacity, and carry lower workloads.
For intra-degree replication, the source is selected from lightly loaded degree-$C$ workers holding $z$, whereas inter-degree replication may use a source worker at any CP degree that holds $z$.
For a selected reclamation, the manager chooses a replica on a worker with greater cache pressure while minimizing the impact on other retained prefixes.

Replication transfers only the blocks missing from the destination's path from the root to $z$, including any missing parent blocks required to form a complete prefix.
Intra-degree replication transfers these blocks between corresponding CP ranks.
Inter-degree replication transfers compatible KV states from a worker at another degree and repartitions them for the destination worker.
A replica is reclaimed only when it is not in use and its removal does not invalidate descendant prefixes retained at the same worker.
A newly created replica becomes visible to request scheduling only after a cache-state report confirms that all required KV states have been installed, while a reclaimed replica is removed from the holder metadata only after its deletion is confirmed.
When worker reconfiguration changes the worker set, the manager reconciles the affected replica records with the resulting placement.

\section{Implementation}

We implement \ourname{} on top of RTP-LLM \cite{tan2026rtp}, a production-oriented LLM serving framework with P/D disaggregation and prefix caching.
Our implementation adds approximately 8,000 lines of C++, Python, and Java code to the prefill engine, request scheduler, worker-management layer, and prefix-cache subsystem, excluding tests and experiment harnesses.
We do not modify decode batching or decode kernels; decode-side changes are limited to propagating the CP configuration epoch and DP-rank identity.
We also implement an offline profiling workflow that fits the prefill service-time model $\widehat E_C(R,P)$ and measures the SLO-preserving capacity $\kappa_C$ for each evaluated model and CP degree deployed in our testbed.
The resulting model parameters and capacity profiles are loaded by the scheduler and reconfiguration controller before serving.

To support in-place worker reconfiguration, we decouple the active CP group from the model state resident on each GPU rank.
Each prefill rank loads its model weights once and retains them throughout serving.
During initialization, \ourname{} reuses communicators already provided by RTP-LLM whenever their rank membership matches a required CP group and creates only the missing groups in the topology-aware grouping lattice.
These process groups and their collective paths are warmed before serving.
They introduce only communicator state and collective buffers while sharing the model state resident on each rank, enabling reconfiguration without process restarts or weight reloads.
We implement the boundary protocol using a versioned control record that specifies the target CP degree and configuration epoch.
Each rank acknowledges its adopted epoch and CP degree, and the controller exposes the resulting workers only after receiving consistent acknowledgments from all participating ranks.

We implement the global prefix trie as an in-memory Java index in the FlexLB control plane, while physical KV blocks remain managed by each worker's native cache index and LRU policy.
Workers report cache-key changes whenever their local cache state is updated, whereas replication and reclamation decisions are made periodically from the maintained trie statistics.
Selected replication operations are issued asynchronously through RTP-LLM's existing KV-transfer path.
During a transfer, the source pins the referenced KV blocks against local eviction, while the destination reserves physical blocks for the missing portion of the selected prefix path.
The destination publishes the resulting cache-key mappings only after the required blocks have been installed across its participating ranks.
Reclamation is implemented through version-checked exact-key deletion, allowing the manager to remove a selected redundant replica while leaving ordinary local eviction to the native LRU policy.
The controller's observation-window duration and the cache manager's scan interval are independently configurable.
The cache manager also exposes a configurable decay factor, per-list action limits, and background-transfer concurrency.
In our experiments, both the controller observation window and the cache-manager scan interval are set to 20 seconds, and the cache dispatcher launches at most one active transfer per scan.

\section{Evaluation}

\subsection{Experimental Setup}

\begin{table}[t]
\centering
\caption{Characteristics of the evaluated workloads.}
\vspace{-0.05in}
\label{tab:workloads}
\small
\setlength{\tabcolsep}{2.5pt}
\begin{tabular}{@{}ccccccc@{}}
\toprule
& \multirow{2}{*}{Reqs.}
& \multicolumn{2}{c}{Input Tokens}
& \multicolumn{2}{c}{Output Tokens}
& \multirow{2}{*}{Reuse} \\
\cmidrule(lr){3-4}
\cmidrule(lr){5-6}
Workload & & Mean & P99 & Mean & P99 & \\
\midrule
L-Eval     & 2,369 & 13,728 & 68,056 & 42.76 & 459 & 79.61\% \\
Tool-Agent & 23,608 & 8,596 & 61,671 & 182 & 898 & 58.56\% \\
Production & 22,446 & 6,952 & 81,724 & 411 & 4,227 & 24.75\% \\
\bottomrule
\end{tabular}
\vspace{-0.1in}
\end{table}

\noindent\textbf{Testbed:}
We evaluate \ourname{} on a cluster of 64 NVIDIA H20-3e GPUs organized into eight nodes, each equipped with eight GPUs, 128 CPU cores, and 960 GB of host memory.
GPUs within each node are connected by NVLink, while the nodes are connected through an RDMA-capable network.
We allocate 40 GPUs to the prefill pool and the remaining 24 GPUs to the independently provisioned decode pool.
We evaluate DeepSeek-V4-Flash-FP8 \cite{xu2026deepseek} and Qwen3-30B-A3B \cite{yang2025qwen3}.
For heterogeneous worker configurations, we use CP degrees $\mathcal C=\{4,8\}$ for DeepSeek-V4-Flash-FP8 and $\mathcal C=\{2,4\}$ for Qwen3-30B-A3B.
For each model, all comparable methods use the same model configuration, 40-GPU prefill budget, 24-GPU decode budget, and aggregate KV-cache capacity.
Across all methods, the decode pool uses the same DP-8 worker configuration and is provisioned so that it does not become the performance bottleneck.

\noindent\textbf{Workloads:}
We evaluate two public workloads, L-Eval \cite{an2024eval} and Mooncake Tool-Agent \cite{toolagent}, together with an anonymized workload trace collected from our production LLM services.
L-Eval covers diverse long-context tasks, including question answering and information retrieval, with widely varying input lengths.
Mooncake Tool-Agent represents agentic serving workloads characterized by recurring system and tool prefixes.
The Production workload is sampled from a recent, anonymized trace collected from our large-scale LLM service.
L-Eval does not provide request-arrival timestamps, so we generate its arrivals using a Poisson process at controlled request rates.
For Mooncake Tool-Agent and the Production workload, we preserve the original request order and timestamps and scale their inter-arrival times to obtain different load levels.
\autoref{tab:workloads} summarizes the request count, request-length distribution, and prefix-reuse potential of each workload.\footnote{
We measure prefix-reuse potential as the token-level hit rate of an oracle prefix cache that retains every previously observed prefix, has unlimited capacity, and makes cached prefixes globally accessible.}\footnote{
The full production traffic typically exhibits a token-level prefix-cache hit rate of 35--55\%.
Because our replay contains only a sampled subset of requests, some cross-request prefix-reuse relationships are absent, and the cache statistics measured during replay do not necessarily match the online values.}

\noindent\textbf{Baselines:}
We compare \ourname{} against two production-oriented baselines implemented on the same serving backend.
Homo uses a homogeneous worker composition in which all prefill workers are degree-4 workers.
Static uses the same worker composition as \ourname{} at each request rate.
It routes requests using an input-length threshold profiled offline for each model and worker composition.
Requests longer than the threshold are assigned to workers with the larger CP degree, whereas the remaining requests are assigned to workers with the smaller CP degree.
We also compare against vLLM \cite{kwon2023efficient} and SGLang \cite{zheng2024sglang} using the same 40-GPU prefill budget.
The evaluated version of vLLM does not support combining CP with data parallelism (DP), so we configure it with homogeneous TP-4 workers.
SGLang supports CP serving but does not allow workers with different CP degrees to coexist in the same serving pool, so we configure it with only degree-4 workers.
Specifically, we do not include LoongServe \cite{wu2024loongserve} because its publicly available implementation does not support the MoE model architectures evaluated in this work, including DeepSeek-V4-Flash-FP8 and Qwen3-30B-A3B.
Supporting these models would require substantial model-specific execution and kernel integration rather than a configuration-only change, preventing a faithful comparison on the same testbed.

\begin{table}[t]
\centering
\caption{TTFT deadlines used to evaluate SLO attainment. The two model deployments use different CP-degree ranges and target different throughput levels; we therefore use model-specific TTFT SLOs.}
\vspace{-0.05in}
\label{tab:ttft-slo}
\small
\setlength{\tabcolsep}{5pt}
\begin{tabular}{cccc}
\toprule
& \multicolumn{3}{c}{Input length} \\
\cmidrule(lr){2-4}
Model
& $<8$K
& $8$--$32$K
& $\geq32$K \\
\midrule
DeepSeek-V4-Flash-FP8             & 0.5\,s & 2\,s & 5\,s \\
Qwen3-30B-A3B    & 1\,s   & 5\,s & 10\,s \\
\bottomrule
\end{tabular}
\vspace{-0.1in}
\end{table}

\noindent\textbf{Metrics and methodology:}
Time-to-first-token (TTFT) is measured from request arrival at the scheduler to the generation of its first output token; it includes queuing, scheduling, synchronous prefix-cache operations, prefill execution, KV transfer to the decode pool, and the first decode iteration.
We report mean and P90 TTFT together with token-weighted TTFT SLO attainment, using the model- and input-length-specific TTFT targets listed in \autoref{tab:ttft-slo}.
Specifically, each request is weighted by its input-token count, and SLO attainment is computed as the total weight of requests meeting the configured TTFT target divided by the total weight of all requests.
For cache experiments, we additionally report the token-level prefix-cache hit rate, defined as the fraction of input tokens reused from cached prefixes.
Before collecting measurements, we warm up each system until its request processing and prefix-cache state stabilize.
We repeat each experiment three times and report the average across these runs.

\begin{figure*}[!ht]
	\centering
    \includegraphics[height=0.4cm]{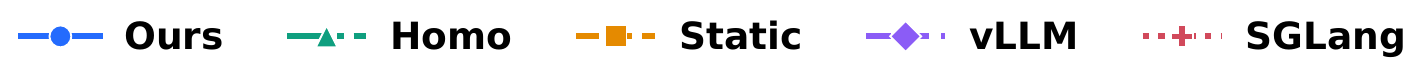} \\
    \vspace{-0.1in}
    \subfigure[L-Eval]{
	\begin{minipage}{0.33\textwidth}{
			\includegraphics[width=1\textwidth]{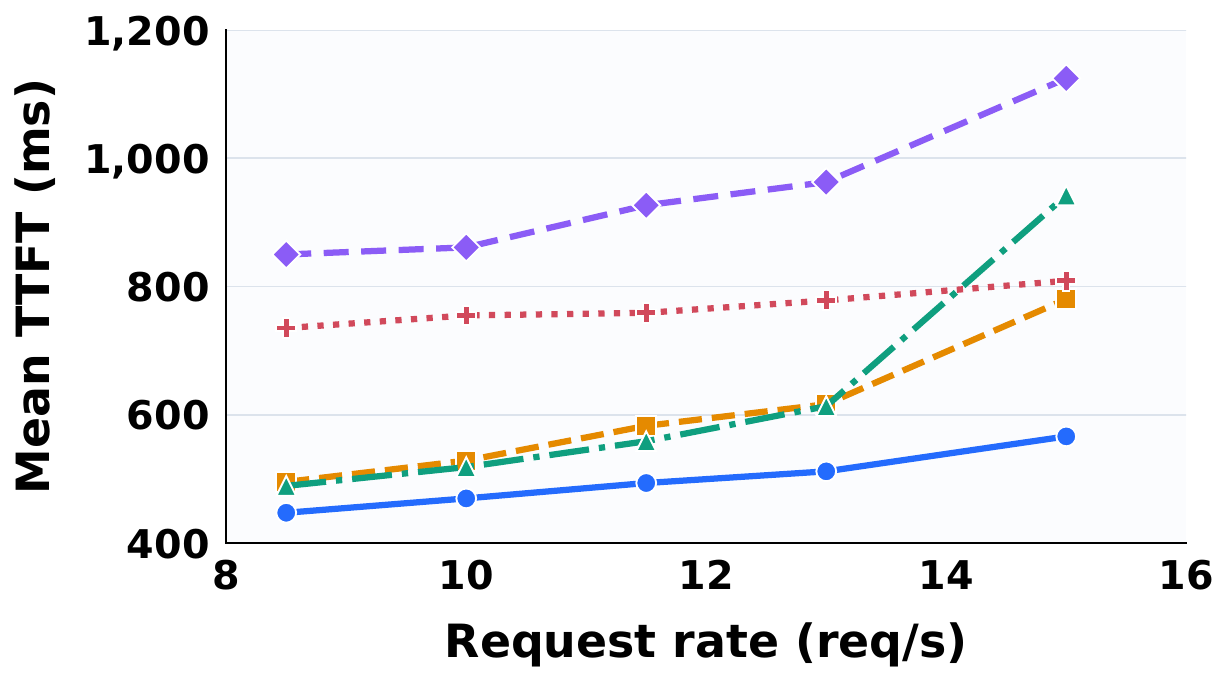}
            \vspace{-0.1in}
		}
	\end{minipage}}%
    \subfigure[Tool-Agent]{
	\begin{minipage}{0.33\textwidth}{
			\includegraphics[width=1\textwidth]{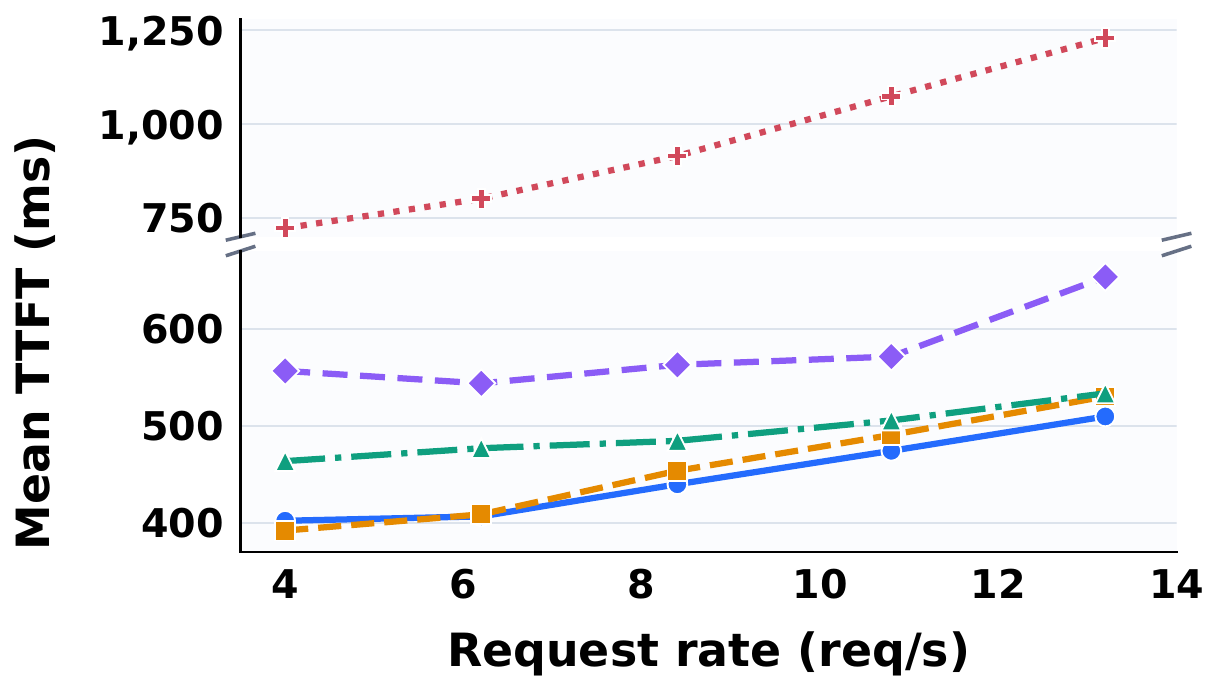}
            \vspace{-0.1in}
		}
	\end{minipage}}%
    \subfigure[Production]{
	\begin{minipage}{0.33\textwidth}{
			\includegraphics[width=1\textwidth]{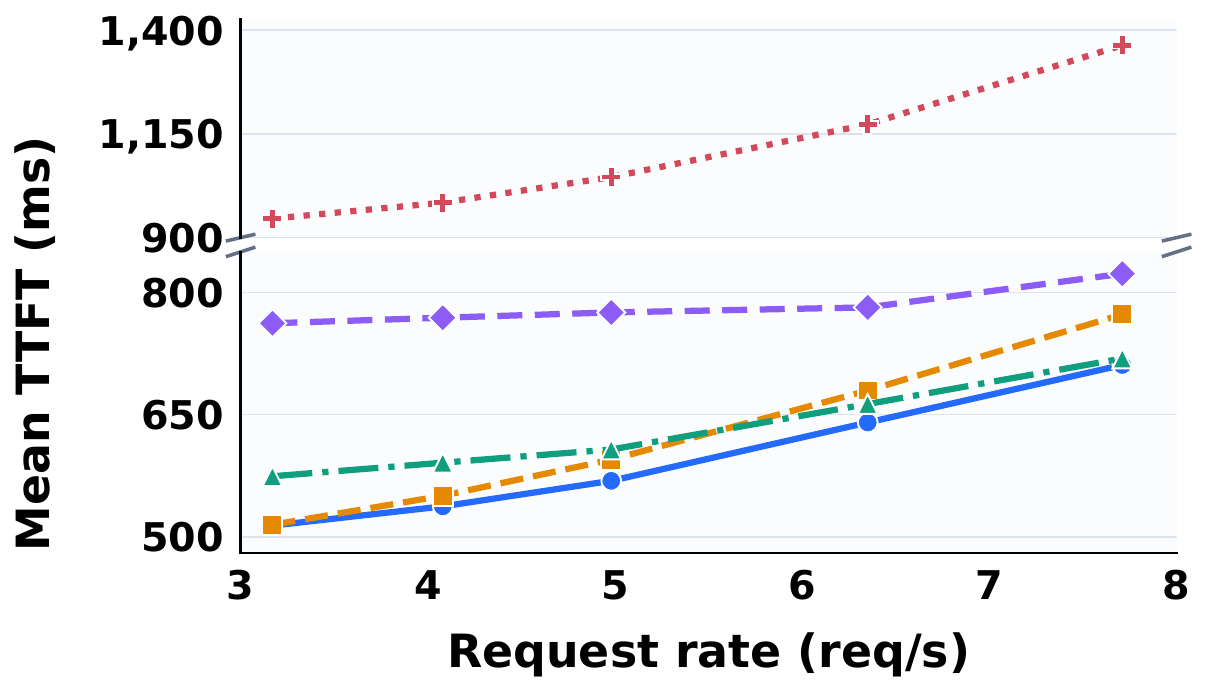}
            \vspace{-0.1in}
		}
	\end{minipage}}%
    \vspace{-0.15in}
    \caption{Mean TTFT for DeepSeek-V4-Flash-FP8 across different workloads.}
    \vspace{-0.15in}
    \label{fig:dsv4-mean}
\end{figure*}

\begin{figure*}[!ht]
	\centering
    \includegraphics[height=0.4cm]{Exp/legend.pdf} \\
    \vspace{-0.1in}
    \subfigure[L-Eval]{
	\begin{minipage}{0.33\textwidth}{
			\includegraphics[width=1\textwidth]{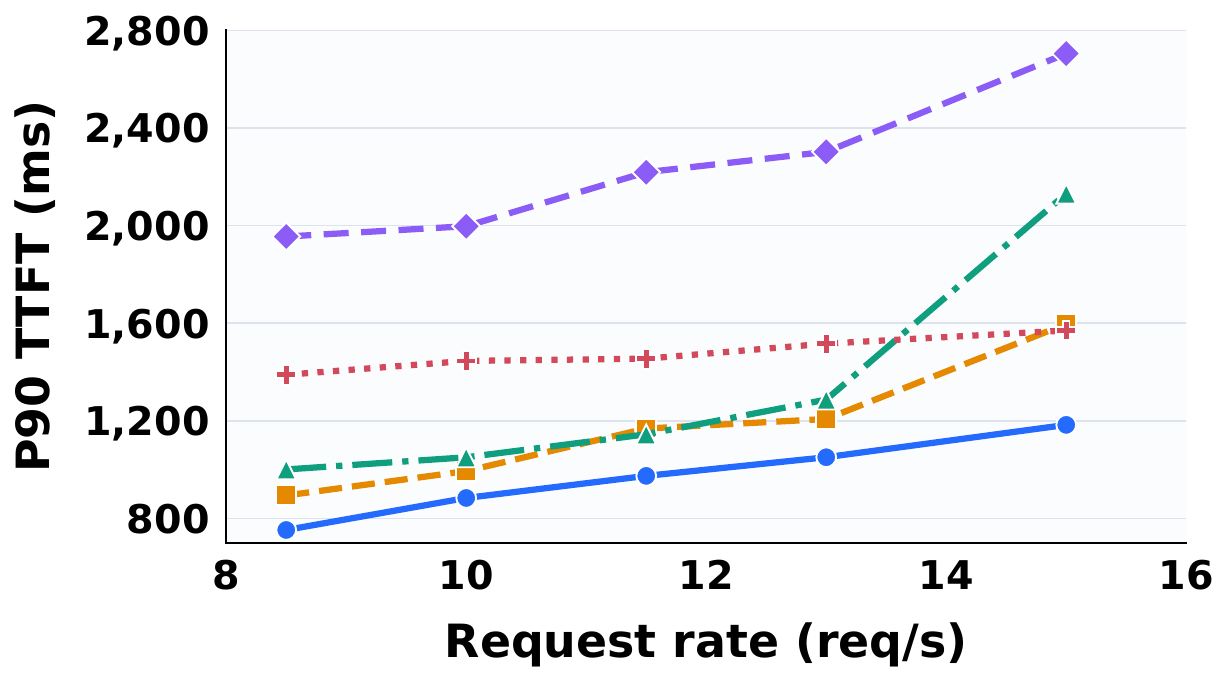}
            \vspace{-0.1in}
		}
	\end{minipage}}%
    \subfigure[Tool-Agent]{
	\begin{minipage}{0.33\textwidth}{
			\includegraphics[width=1\textwidth]{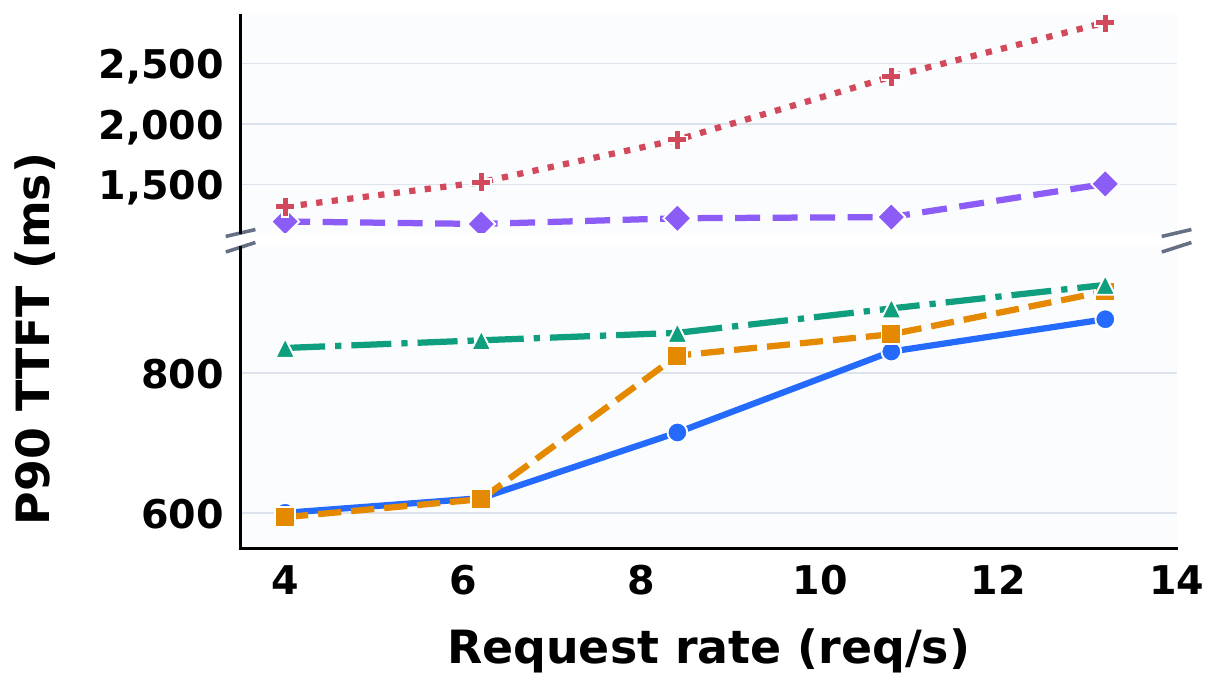}
            \vspace{-0.1in}
		}
	\end{minipage}}%
    \subfigure[Production]{
	\begin{minipage}{0.33\textwidth}{
			\includegraphics[width=1\textwidth]{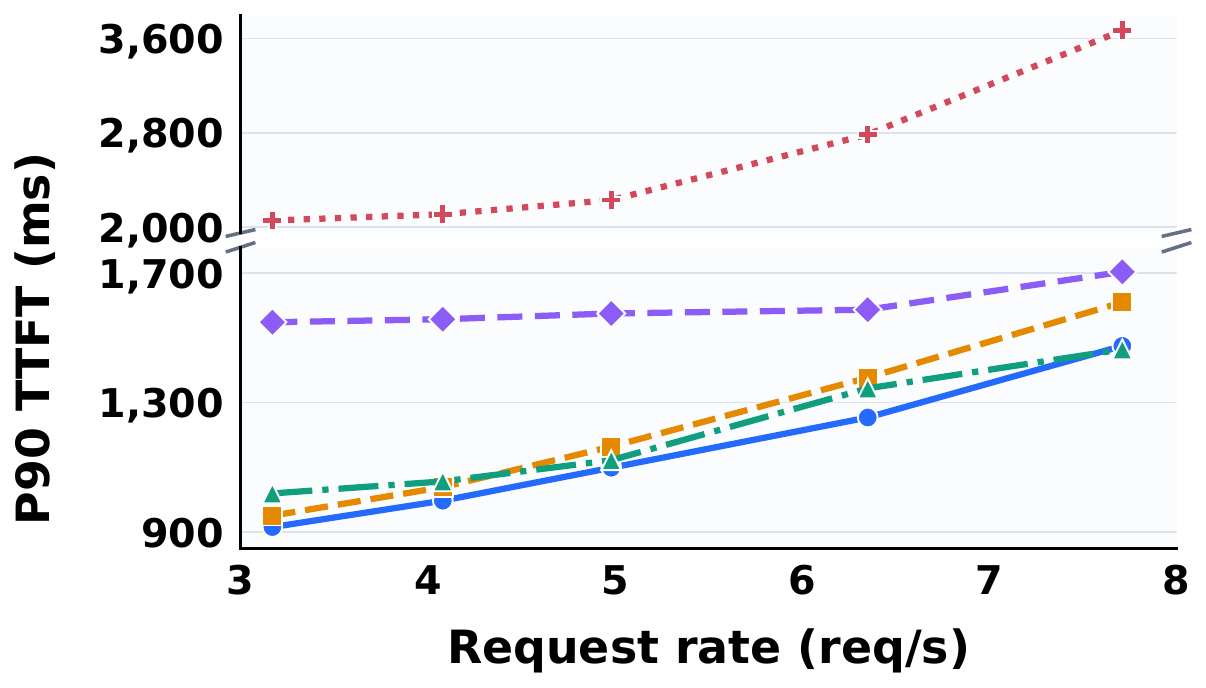}
            \vspace{-0.1in}
		}
	\end{minipage}}%
    \vspace{-0.15in}
    \caption{P90 TTFT for DeepSeek-V4-Flash-FP8 across different workloads.}
    \vspace{-0.15in}
    \label{fig:dsv4-p90}
\end{figure*}

\begin{figure*}[!ht]
	\centering
    \includegraphics[height=0.4cm]{Exp/legend.pdf} \\
    \vspace{-0.1in}
    \subfigure[L-Eval]{
	\begin{minipage}{0.33\textwidth}{
			\includegraphics[width=1\textwidth]{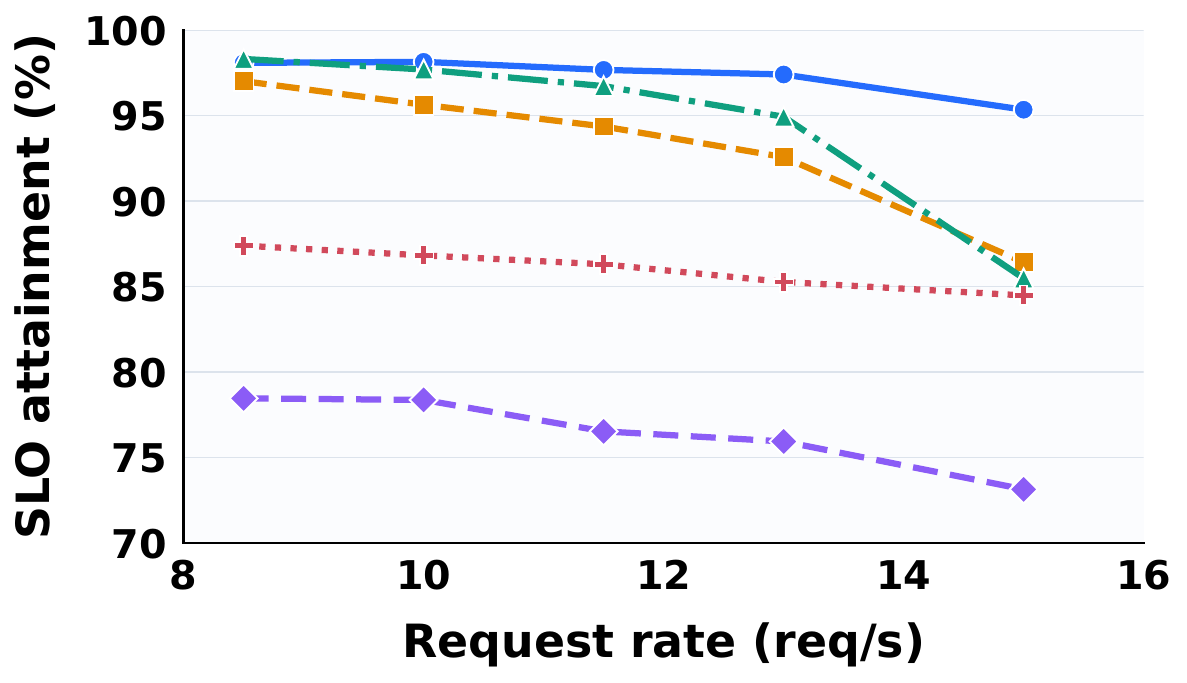}
            \vspace{-0.1in}
		}
	\end{minipage}}%
    \subfigure[Tool-Agent]{
	\begin{minipage}{0.33\textwidth}{
			\includegraphics[width=1\textwidth]{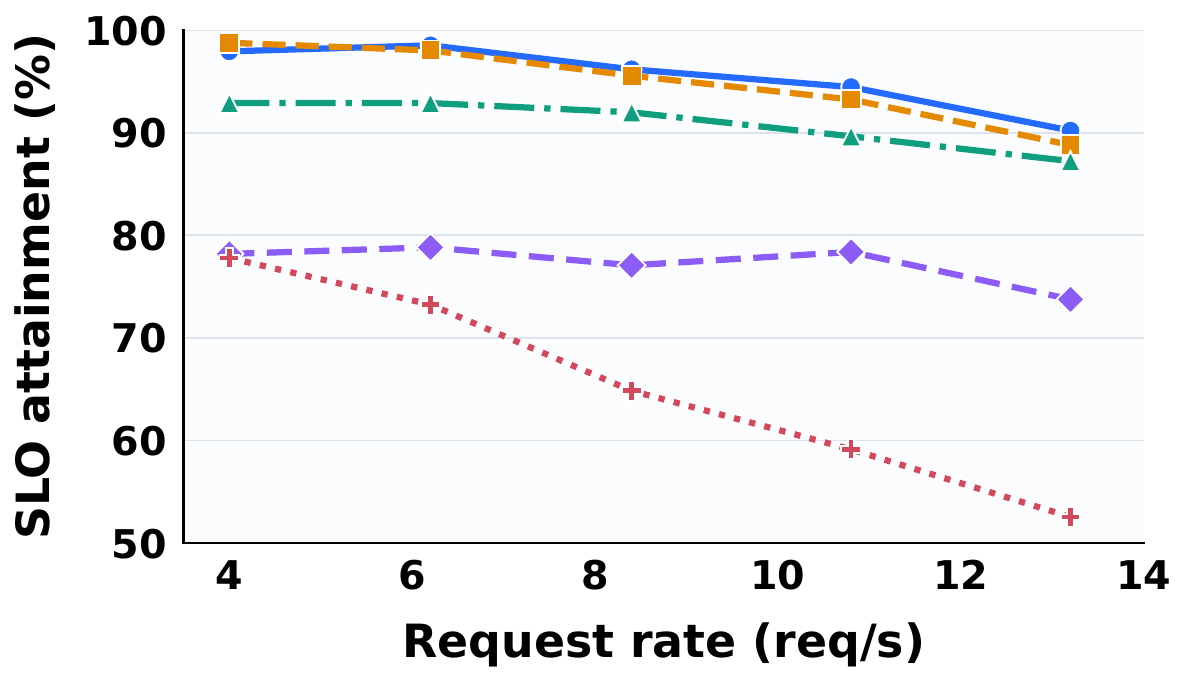}
            \vspace{-0.1in}
		}
	\end{minipage}}%
    \subfigure[Production]{
	\begin{minipage}{0.33\textwidth}{
			\includegraphics[width=1\textwidth]{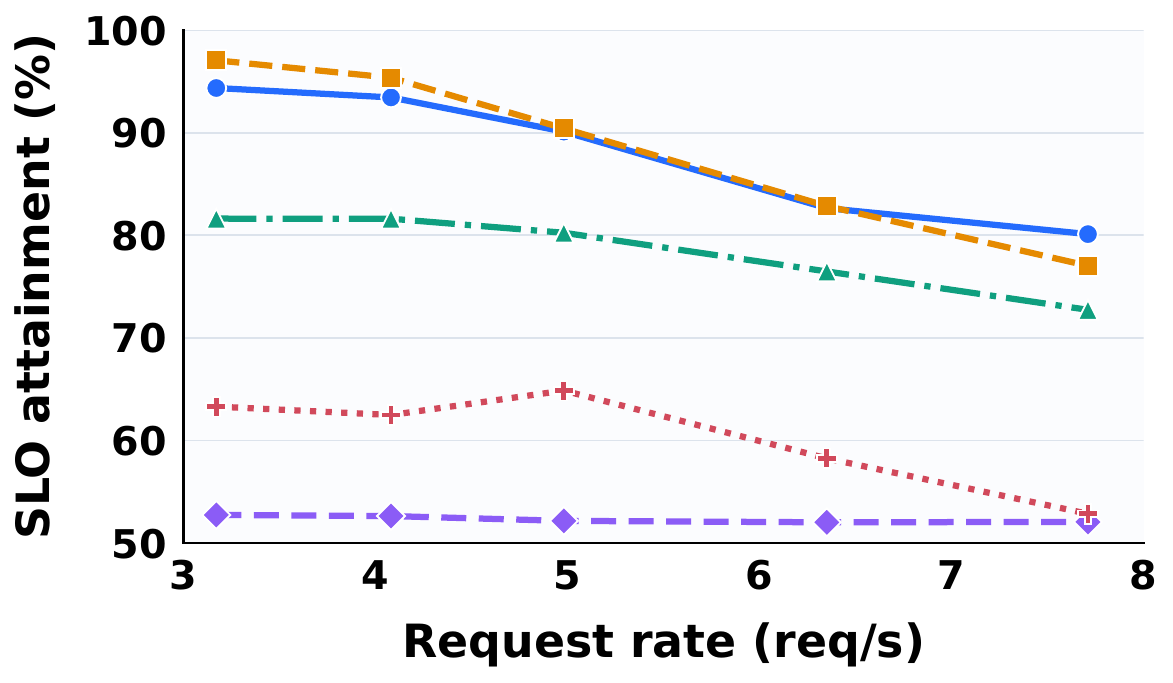}
            \vspace{-0.1in}
		}
	\end{minipage}}%
    \vspace{-0.15in}
    \caption{SLO attainment for DeepSeek-V4-Flash-FP8 across different workloads.}
    \vspace{-0.15in}
    \label{fig:dsv4-slo}
\end{figure*}

\subsection{End-to-End Serving Performance}

\begin{figure*}[!ht]
	\centering
    \includegraphics[height=0.4cm]{Exp/legend.pdf} \\
    \vspace{-0.1in}
    \subfigure[L-Eval]{
	\begin{minipage}{0.33\textwidth}{
			\includegraphics[width=1\textwidth]{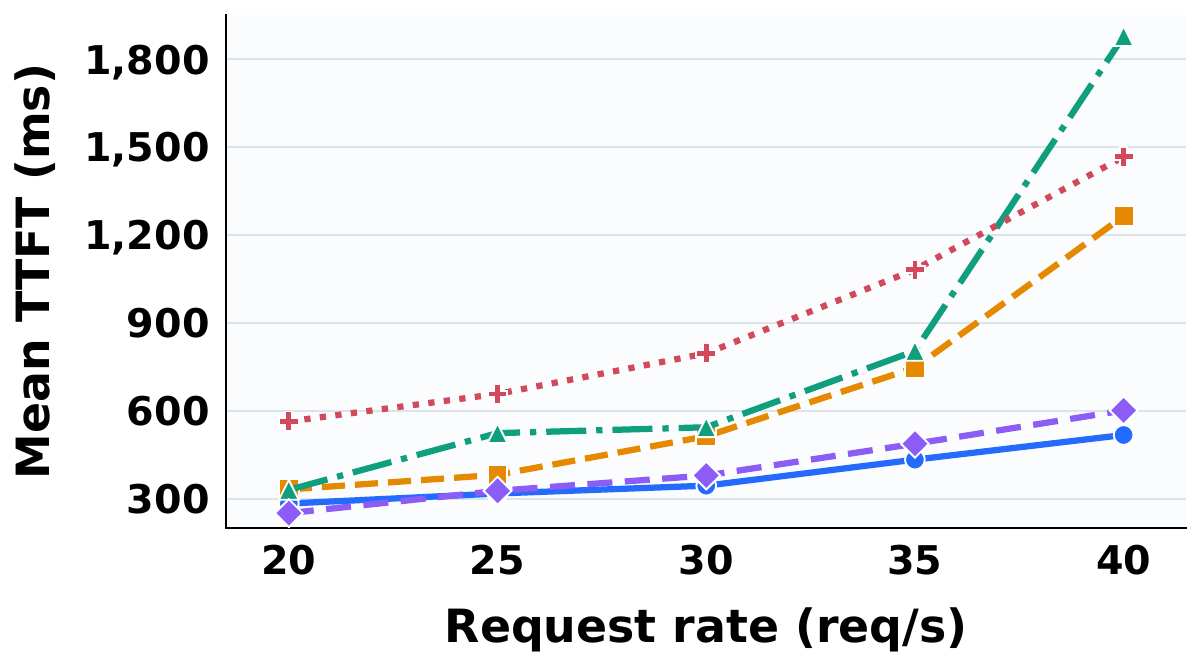}
            \vspace{-0.1in}
		}
	\end{minipage}}%
    \subfigure[Tool-Agent]{
	\begin{minipage}{0.33\textwidth}{
			\includegraphics[width=1\textwidth]{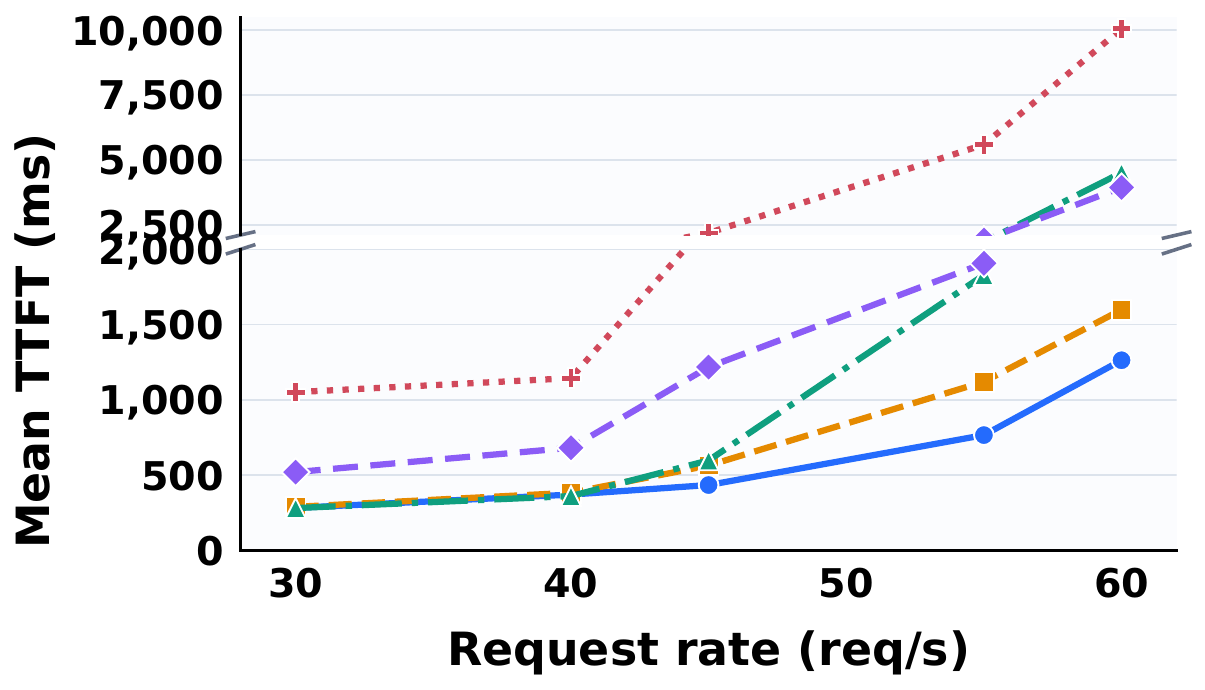}
            \vspace{-0.1in}
		}
	\end{minipage}}%
    \subfigure[Production]{
	\begin{minipage}{0.33\textwidth}{
			\includegraphics[width=1\textwidth]{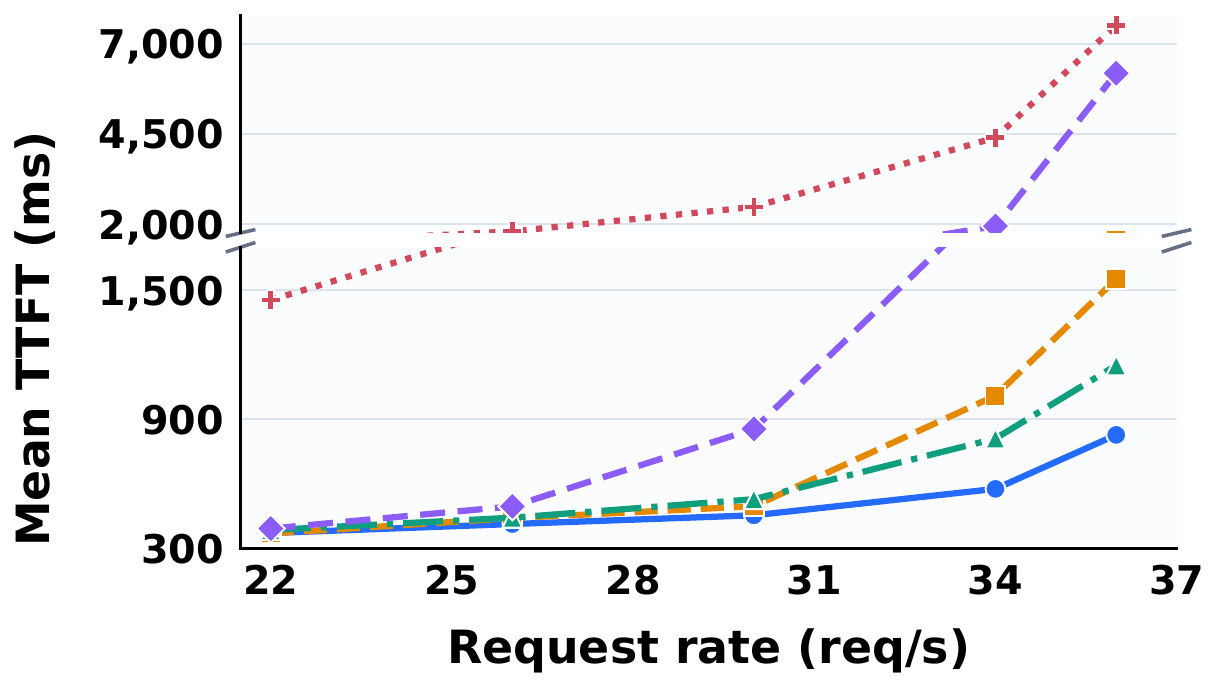}
            \vspace{-0.1in}
		}
	\end{minipage}}%
    \vspace{-0.15in}
    \caption{Mean TTFT for Qwen3-30B-A3B across different workloads.}
    \vspace{-0.15in}
    \label{fig:qwen-mean}
\end{figure*}

\begin{figure*}[!ht]
	\centering
    \includegraphics[height=0.4cm]{Exp/legend.pdf} \\
    \vspace{-0.1in}
    \subfigure[L-Eval]{
	\begin{minipage}{0.33\textwidth}{
			\includegraphics[width=1\textwidth]{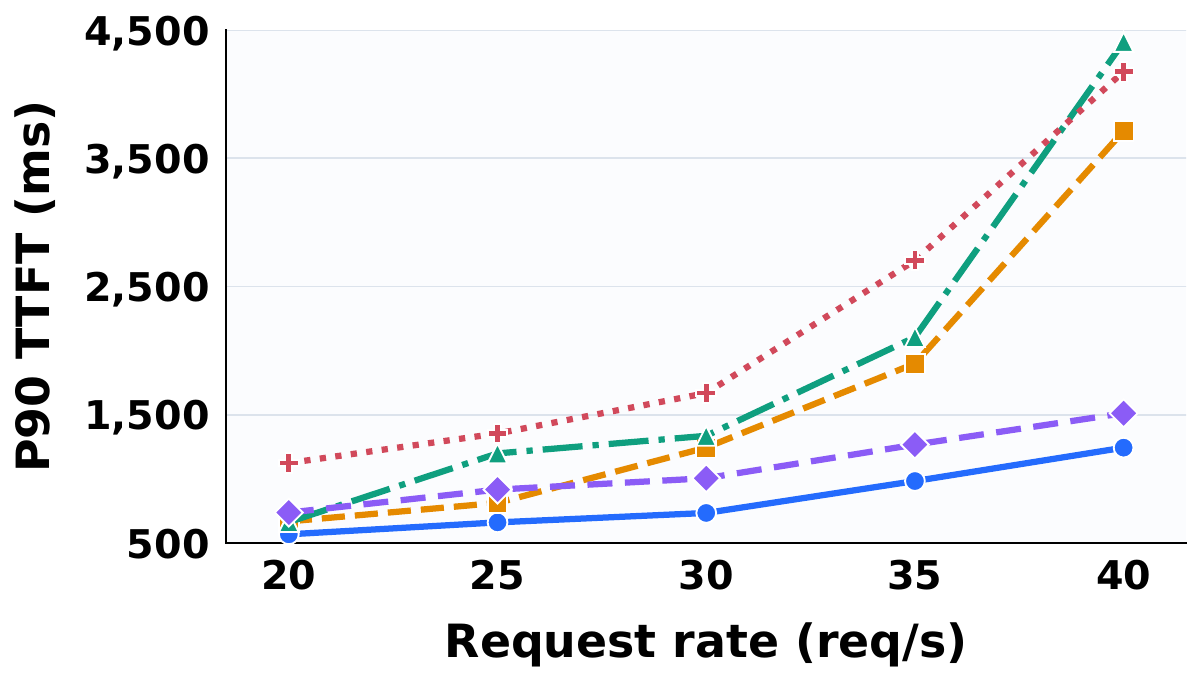}
            \vspace{-0.1in}
		}
	\end{minipage}}%
    \subfigure[Tool-Agent]{
	\begin{minipage}{0.33\textwidth}{
			\includegraphics[width=1\textwidth]{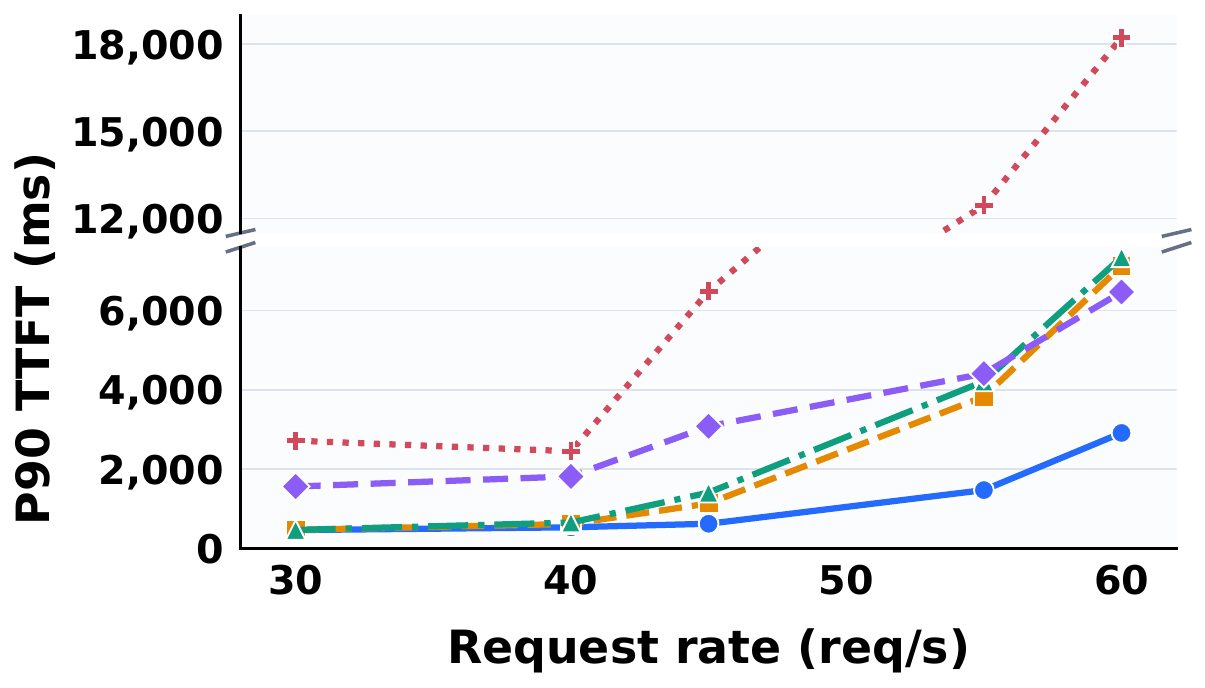}
            \vspace{-0.1in}
		}
	\end{minipage}}%
    \subfigure[Production]{
	\begin{minipage}{0.33\textwidth}{
			\includegraphics[width=1\textwidth]{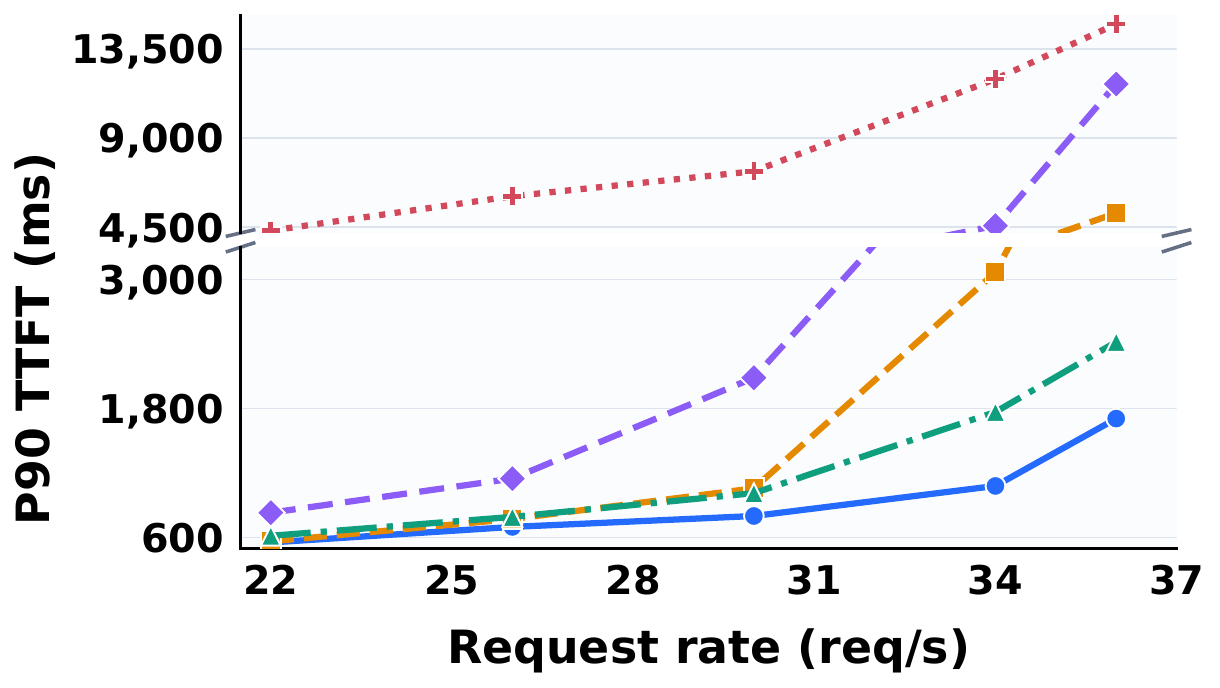}
            \vspace{-0.1in}
		}
	\end{minipage}}%
    \vspace{-0.15in}
    \caption{P90 TTFT for Qwen3-30B-A3B across different workloads.}
    \vspace{-0.15in}
    \label{fig:qwen-p90}
\end{figure*}

\begin{figure*}[!ht]
	\centering
    \includegraphics[height=0.4cm]{Exp/legend.pdf} \\
    \vspace{-0.1in}
    \subfigure[L-Eval]{
	\begin{minipage}{0.33\textwidth}{
			\includegraphics[width=1\textwidth]{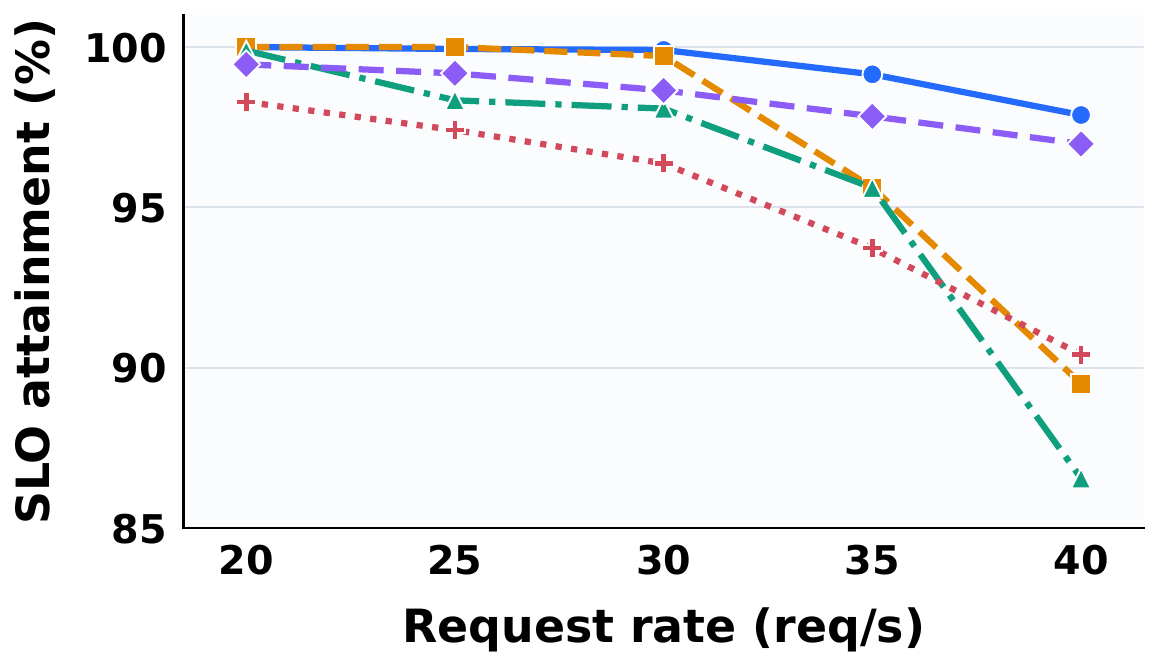}
            \vspace{-0.1in}
		}
	\end{minipage}}%
    \subfigure[Tool-Agent]{
	\begin{minipage}{0.33\textwidth}{
			\includegraphics[width=1\textwidth]{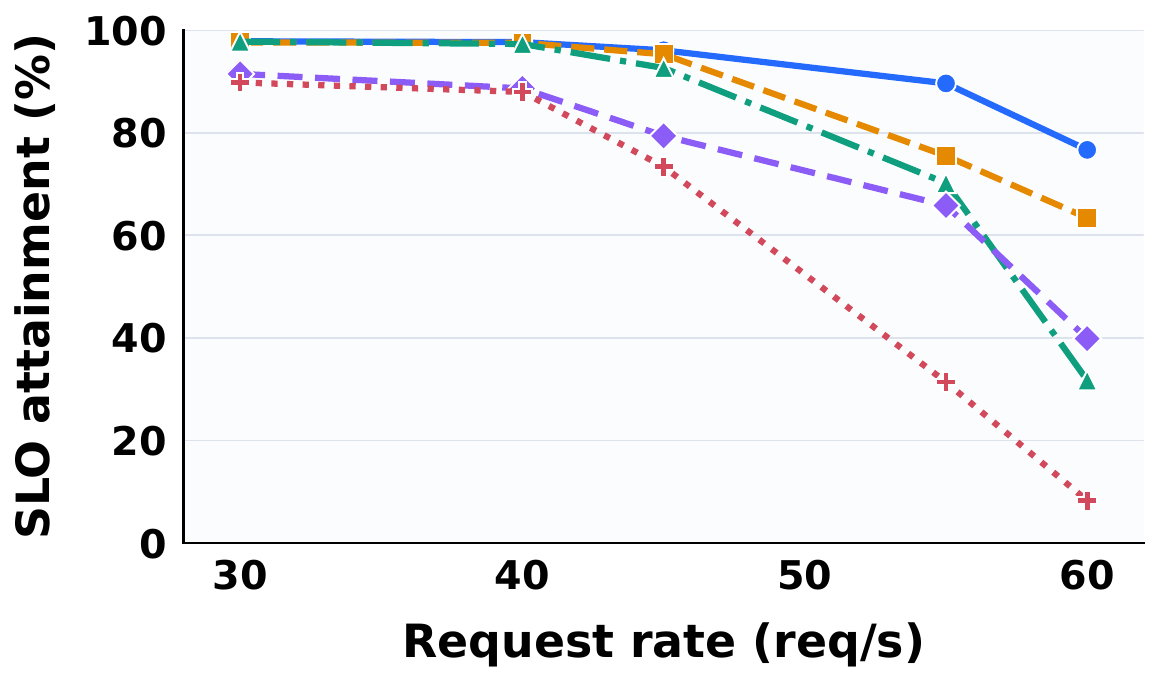}
            \vspace{-0.1in}
		}
	\end{minipage}}%
    \subfigure[Production]{
	\begin{minipage}{0.33\textwidth}{
			\includegraphics[width=1\textwidth]{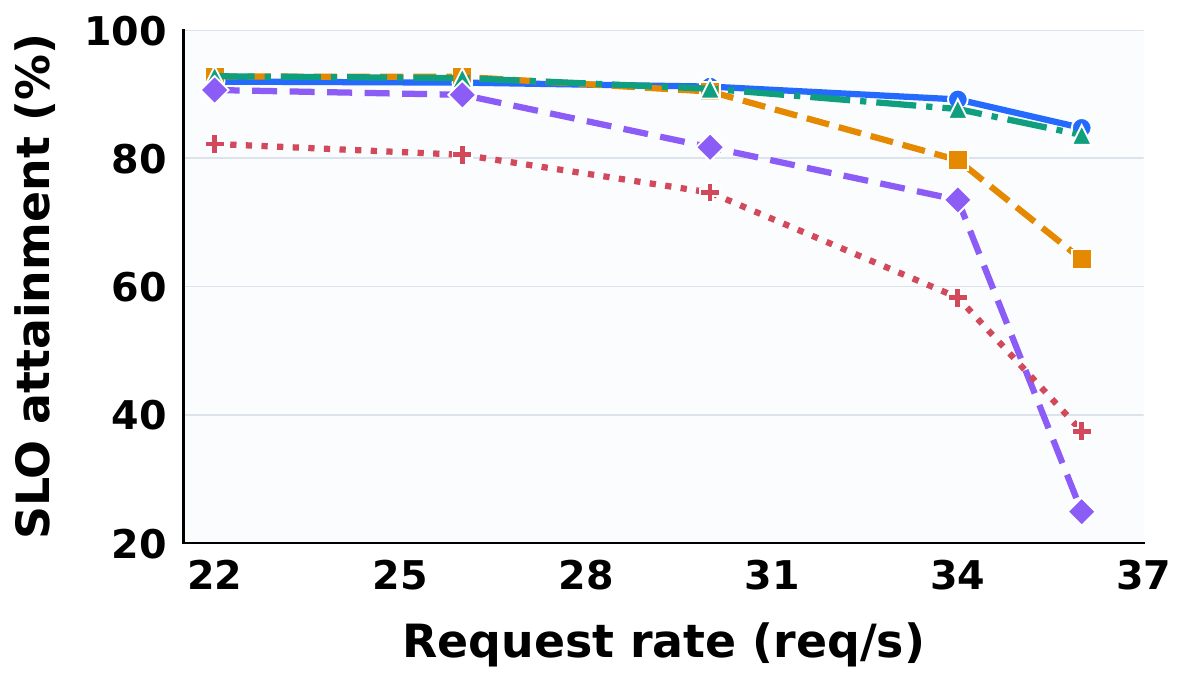}
            \vspace{-0.1in}
		}
	\end{minipage}}%
    \vspace{-0.15in}
    \caption{SLO attainment for Qwen3-30B-A3B across different workloads.}
    \vspace{-0.15in}
    \label{fig:qwen-slo}
\end{figure*}

Figures~\ref{fig:dsv4-mean}--\ref{fig:qwen-slo} present the end-to-end results across two models, three workloads, and different request rates.
Overall, \ourname{} achieves the lowest or comparable mean and P90 TTFT and the highest or near-highest SLO attainment, with its advantages becoming clearer as the baselines accumulate delays under high load.

\noindent\textbf{DeepSeek-V4-Flash-FP8:}
\ourname{} generally matches or outperforms Homo and Static while providing lower mean and P90 TTFT than vLLM and SGLang.
Static may assign prefix-sharing requests with different input lengths to workers with different CP degrees, while Homo's single CP degree cannot balance concurrency for short requests and parallelism for long requests.
In contrast, \ourname{} jointly considers worker load, worker-specific prefix reuse, and predicted execution time at each CP degree.
At the highest request rates, it reduces mean TTFT by 27.4\%, 3.9\%, and 1.1\% on L-Eval, Tool-Agent, and Production, respectively, compared with the strongest baseline.
It also reduces P90 TTFT by 24.6\% and 4.3\% on L-Eval and Tool-Agent while remaining within 0.7\% of Homo on Production, with corresponding SLO attainment improvements of 8.9, 1.4, and 3.1 percentage points.

\noindent\textbf{Qwen3-30B-A3B:}
Individual baselines may match \ourname{} at lower request rates, but their latency increases more rapidly as the load grows, particularly on Tool-Agent and Production.
At the highest request rates, \ourname{} appropriately places requests across degree-2 and degree-4 workers, reducing mean TTFT by 13.9\%, 21.0\%, and 28.1\% on L-Eval, Tool-Agent, and Production, respectively, relative to the strongest baseline.
The corresponding P90 reductions are 17.8\%, 55.0\%, and 29.5\%, while SLO attainment improves by 0.9, 13.3, and 1.1 percentage points.
These tail-latency improvements show that \ourname{} effectively slows queue buildup as the request rate increases.

\subsection{Evaluation of Worker Reconfiguration}

\begin{figure*}[!ht]
	\centering
    \subfigure[Load adaptation]{
	\begin{minipage}{0.33\textwidth}{
			\includegraphics[width=1\textwidth]{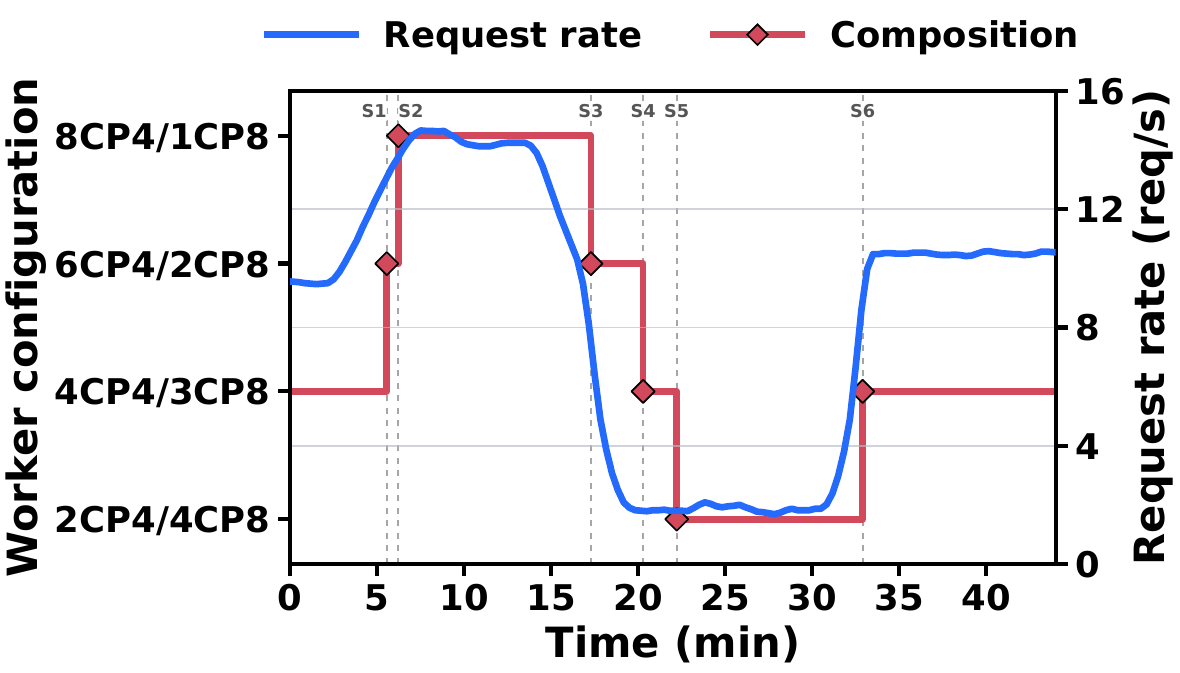}
            \vspace{-0.1in}
		}
	\end{minipage}}%
    \subfigure[Mean TTFT comparison]{
	\begin{minipage}{0.33\textwidth}{
			\includegraphics[width=1\textwidth]{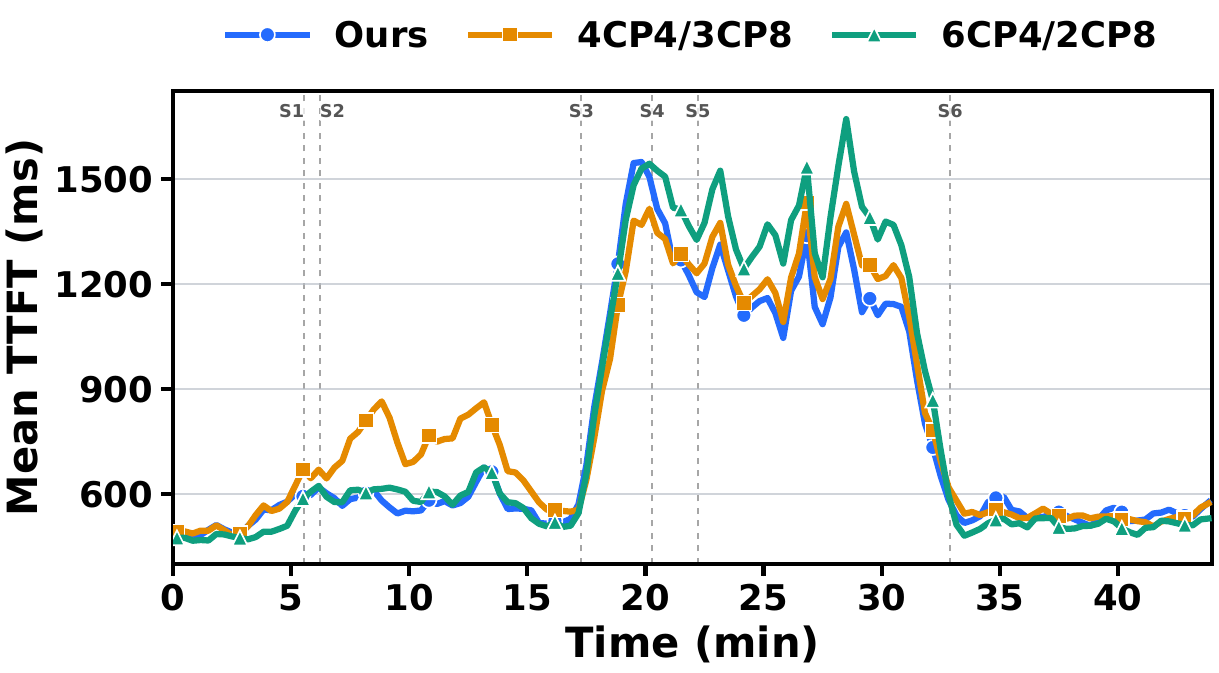}
            \vspace{-0.1in}
		}
	\end{minipage}}%
    \subfigure[Latency breakdown]{
	\begin{minipage}{0.33\textwidth}{
			\includegraphics[width=1\textwidth]{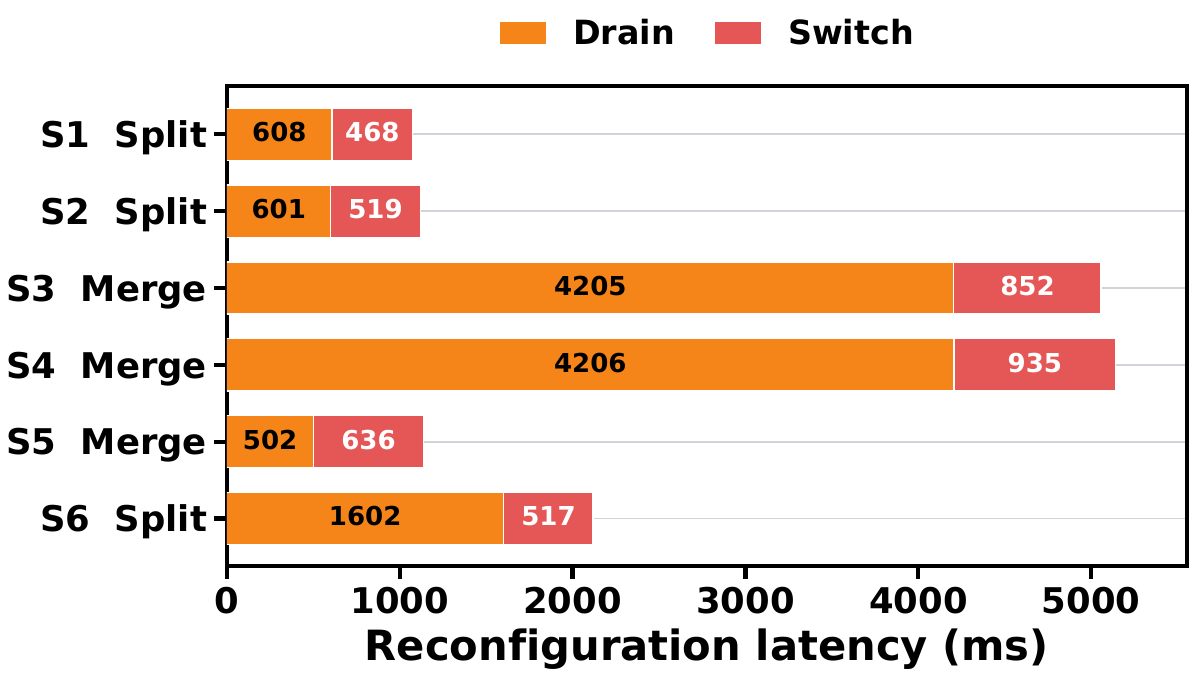}
            \vspace{-0.1in}
		}
	\end{minipage}}%
    \vspace{-0.15in}
    \caption{Experiments on worker reconfiguration.}
    \vspace{-0.2in}
    \label{fig:cp-switch}
\end{figure*}

We evaluate DeepSeek-V4-Flash-FP8 on a time-varying workload derived from Mooncake Tool-Agent, alternating between high-rate phases dominated by short requests and low-rate phases with more long requests.
We compare \ourname{} against two fixed worker-composition baselines, $4\mathrm{CP}4/3\mathrm{CP}8$ and $6\mathrm{CP}4/2\mathrm{CP}8$.
The results are shown in \autoref{fig:cp-switch}.

We find that no single fixed composition performs best throughout the workload:
the $4\mathrm{CP}4/3\mathrm{CP}8$ baseline provides insufficient concurrency when short requests dominate, whereas $6\mathrm{CP}4/2\mathrm{CP}8$ provides too few large-degree workers when the fraction of long requests increases.
By adjusting its worker composition shortly after each sustained workload change, \ourname{} matches or outperforms the better fixed baseline in each phase.
When the request rate is moderate and stable, its mean TTFT is comparable to both fixed baselines.
As the request rate increases between 6 and 17 minutes, \ourname{} switches to $8\mathrm{CP}4/1\mathrm{CP}8$ to increase serving concurrency, reducing mean TTFT by 19.6\% compared with $4\mathrm{CP}4/3\mathrm{CP}8$.
When the request rate subsequently decreases and long requests become more common between 22 and 33 minutes, \ourname{} switches to $2\mathrm{CP}4/4\mathrm{CP}8$ to provide more large-degree workers, reducing mean TTFT by 5.5\% and 14.5\% compared with $4\mathrm{CP}4/3\mathrm{CP}8$ and $6\mathrm{CP}4/2\mathrm{CP}8$, respectively.

To better understand the overhead of worker reconfiguration, we break down its latency into drain and switch stages.
During draining, the selected workers stop accepting new requests, redirect their queued requests to other eligible workers, and allow ongoing prefill requests to finish.
Once draining completes, the participating ranks activate the target pre-created communication groups and expose the resulting workers for scheduling while keeping model weights and retained KV data resident in GPU memory.
As shown in the figure, each split or merge operation takes between 1.1 and 5.1 seconds, with most of the variation coming from the 0.5--4.2-second drain stage while the system waits for ongoing prefill requests to finish.
The switch stage consistently completes within one second.
These results show that \ourname{} can change its worker composition within seconds and respond promptly to sustained workload changes.

\begin{figure*}[!ht]
	\centering
    \subfigure[Mean TTFT]{
	\begin{minipage}{0.33\textwidth}{
			\includegraphics[width=1\textwidth]{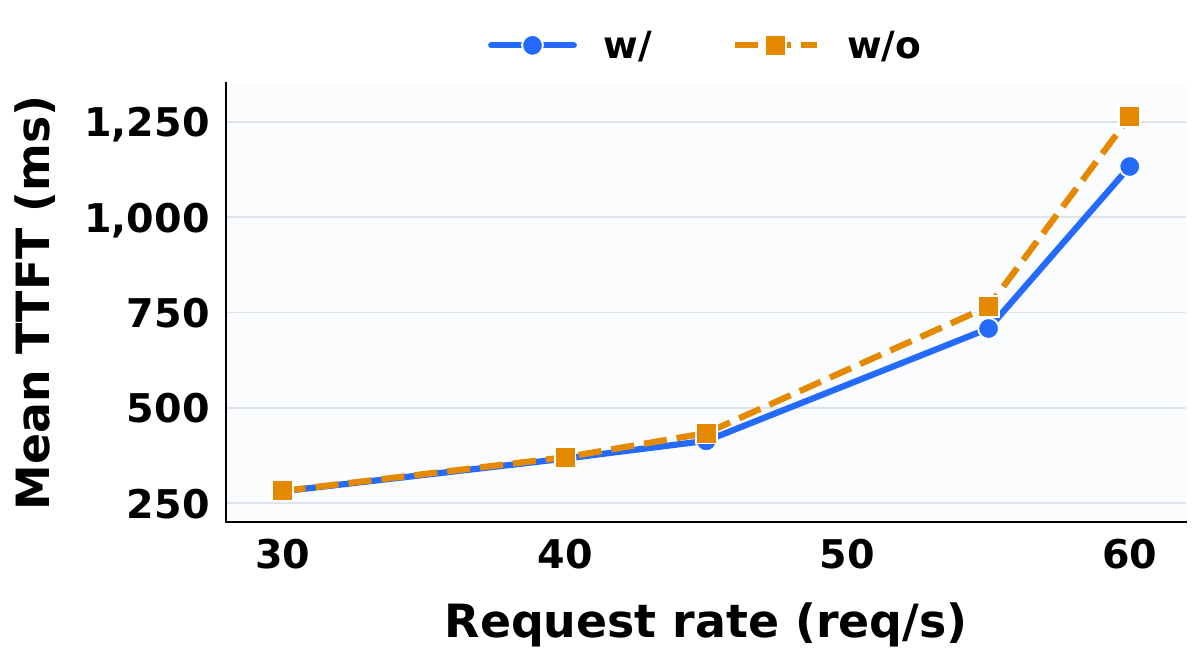}
            \vspace{-0.1in}
		}
	\end{minipage}}%
    \subfigure[P90 TTFT]{
	\begin{minipage}{0.33\textwidth}{
			\includegraphics[width=1\textwidth]{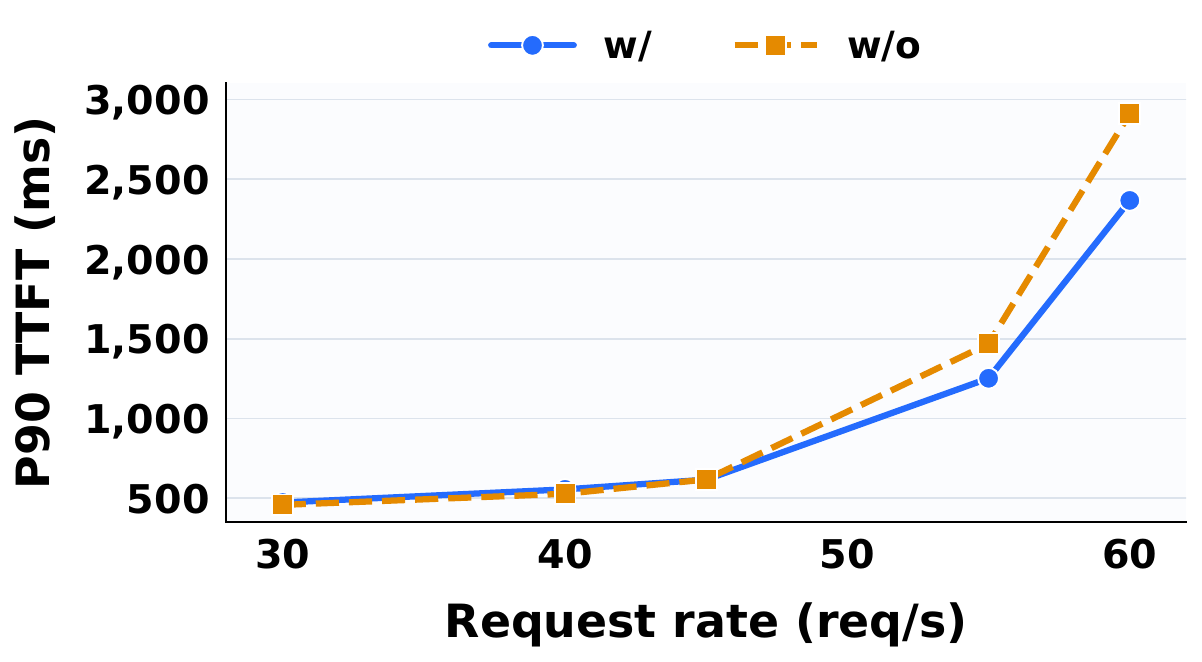}
            \vspace{-0.1in}
		}
	\end{minipage}}%
    \subfigure[Prefix-cache hit rate]{
	\begin{minipage}{0.33\textwidth}{
			\includegraphics[width=1\textwidth]{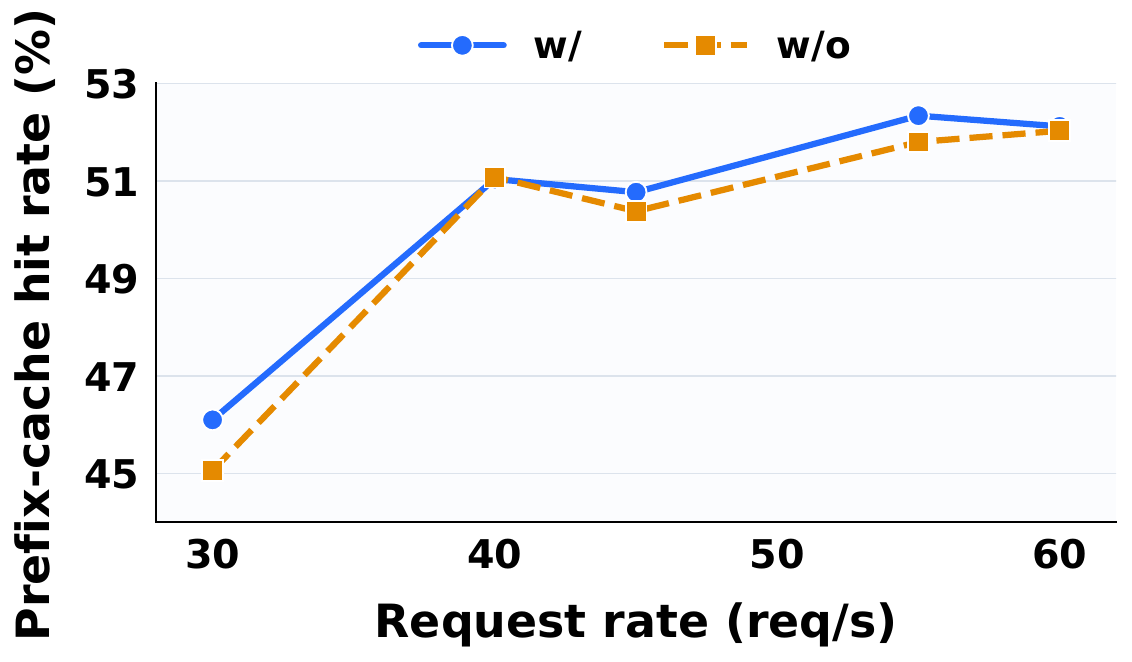}
            \vspace{-0.1in}
		}
	\end{minipage}}%
    \vspace{-0.15in}
    \caption{Experiments on prefix-cache ablation.}
    \vspace{-0.15in}
    \label{fig:cache-ablation}
\end{figure*}

\subsection{Prefix-Cache Ablation}

To demonstrate the effect of the prefix-cache manager, we also perform an ablation study using Qwen3-30B-A3B and the Tool-Agent workload.
For a controlled comparison, we fix the worker composition and compare the full \ourname{} against a variant that retains worker-local prefix caching but disables prefix replication and reclamation.
Both variants use the same request scheduler and per-worker cache capacity. 
In the ablated variant, each worker continues to insert prefixes and handle cache pressure through its native LRU policy. 
We vary the request rate and show mean and P90 TTFT together with the token-level prefix-cache hit rate in \autoref{fig:cache-ablation}.

At lower request rates, the two variants provide similar TTFT because worker queues remain short.
As the request rate increases, however, the prefix-cache manager yields larger latency reductions.
Without replication, preserving a cache hit may restrict a request to the only worker holding its matched prefix, even when that worker is heavily loaded.
Intra-degree replication creates additional copies of frequently accessed prefixes among workers with the same CP degree, allowing the scheduler to select a less-loaded worker without sacrificing prefix reuse.
Inter-degree replication makes a prefix available at a CP degree where it was previously absent, avoiding the need to choose between cache reuse and a CP degree better suited to the request.
The manager also reclaims underused replicas to recover cache capacity for prefixes with greater demand.
At 60 requests/s, the complete \ourname{} reduces mean and P90 TTFT by 10.3\% and 18.7\%, respectively, while modestly improving the token-level prefix-cache hit rate.
These results show that the manager reduces TTFT primarily by expanding cache-preserving placement choices both within and across CP degrees, rather than merely increasing the cache hit rate.

\section{Related Work}
\label{sec:related_work}

\subsection{LLM Serving Systems}
Existing LLM serving systems optimize request execution at different levels.
At the local level, systems such as Orca \cite{yu2022orca}, Sarathi-Serve \cite{agrawal2024taming}, FastServe \cite{wu2026fastserve}, and Medha \cite{agrawal2024medha} schedule requests within a model instance by controlling their batching, execution order, and interleaving.
At the cluster level, systems such as Llumnix \cite{sun2024llumnix}, MuxServe \cite{duan2024muxserve}, and AlpaServe \cite{li2023alpaserve} route or migrate requests and place model instances across available GPUs.
Another line of work, including DistServe \cite{zhong2024distserve}, Splitwise \cite{patel2024splitwise}, and TetriInfer \cite{hu2024inference}, disaggregates prefill and decode into separate worker pools so that the two stages can be scheduled and provisioned independently.
These systems generally perform scheduling over workers whose parallel configurations have already been determined.

\subsection{Dynamic Parallelism for LLM Serving}

Most existing LLM serving systems operate over workers whose data-parallel (DP), tensor-parallel (TP), or pipeline-parallel (PP) configurations are fixed at initialization \cite{lin2026kvdrive, jiang2024chameleon, hong2025sola, gong2025past, yang2025lserve}.
Several recent systems have explored reconfiguring model parallelism at runtime to accommodate changes in workloads or resource availability.
For example, SpotServe \cite{miao2024spotserve}, Gyges \cite{chen2025gyges}, CoCoServe \cite{wu2025unlock}, and OServe \cite{jiang2026oserve} adapt the choice of model-parallel configuration to changes in workload or GPU availability.
A complementary line of work, including PipeLive \cite{bai2026pipelive}, ReMP \cite{yuan2026remp}, and Flying Serving \cite{gao2026flying}, further reduces service interruption and state-transfer overhead when changing parallel configurations online.
These systems primarily focus on limiting service interruption and state-transfer overhead during model-parallel reconfiguration.


A separate line of work explores dynamic context or sequence parallelism for LLM serving.
LoongServe \cite{wu2024loongserve} dynamically forms groups of elastic instances and adjusts the parallelism of active batches across serving iterations.
NanoCP \cite{chen2026nanocp} specifically targets the decode stage of MoE serving: it selects a CP degree for each active request to balance attention and expert execution and reduce TPOT.
Both systems adapt the execution plans of active requests or batches rather than maintaining a persistent worker composition across workload windows.
In contrast, \ourname{} targets prefill serving: it organizes a fixed GPU budget into persistent CP workers with different degrees, routes incoming requests among them, and reconfigures their composition as aggregate demand evolves.
This persistent organization also provides stable units for request queuing and prefix-cache placement across requests.

\subsection{Prefix Caching and Cache Management}

Prior work on prefix caching reuses shared prompt states across requests to avoid redundant prefill computation, as demonstrated by Prompt Cache \cite{gim2023prompt}, SGLang \cite{zheng2024sglang}, and ChunkAttention \cite{ye2024chunkattention}.
Cache-aware serving systems such as Preble \cite{srivatsa2025preble}, AlignedServe \cite{bai2026alignedserve}, and Apt-Serve \cite{gao2025apt} coordinate cache reuse with request scheduling or batching.
Another line of work, including Mooncake \cite{qin2024mooncake}, MemServe \cite{hu2024memserve}, LMCache \cite{liu2025lmcache}, Infinite-LLM \cite{lin2024infinite}, and KVFlow \cite{pan2026kvflow}, manages KV states across workers and memory tiers.
TokenLake \cite{wu2025tokenlake} further constructs a unified segment-level prefix-cache pool to support fine-grained elastic long-context serving.

\section{Conclusion}

In this paper, we presented \ourname{}, an LLM inference system that adaptively manages context parallelism under heterogeneous and changing production workloads.
\ourname{} supports persistent workers with different CP degrees and provides a request-scheduling policy for coordinating request placement.
To adapt at the cluster timescale, \ourname{} dynamically adjusts the worker composition through lightweight split and merge operations.
It further provides a global prefix-cache manager that tracks prefix demand and dynamically replicates and reclaims cached prefixes.
Experiments with public and production workloads demonstrate the effectiveness of \ourname{} under diverse workload conditions.

\newpage
\clearpage
{
\balance
\bibliographystyle{unsrt}
\bibliography{reference}

@article{xu2026deepseek,
  title={Deepseek-v4: Towards highly efficient million-token context intelligence},
  author={Xu, Anyi and Lin, Bangcai and Xue, Bing and Wang, Bingxuan and Xu, Bingzheng and Wu, Bochao and Zhang, Bowei and Lin, Chaofan and Dong, Chen and Ling, Chenchen and others},
  journal={arXiv preprint arXiv:2606.19348},
  year={2026}
}

@article{yang2025qwen3,
  title={Qwen3 technical report},
  author={Yang, An and Li, Anfeng and Yang, Baosong and Zhang, Beichen and Hui, Binyuan and Zheng, Bo and Yu, Bowen and Gao, Chang and Huang, Chengen and Lv, Chenxu and others},
  journal={arXiv preprint arXiv:2505.09388},
  year={2025}
}

@article{achiam2023gpt,
  title={Gpt-4 technical report},
  author={Achiam, Josh and Adler, Steven and Agarwal, Sandhini and Ahmad, Lama and Akkaya, Ilge and Aleman, Florencia Leoni and Almeida, Diogo and Altenschmidt, Janko and Altman, Sam and Anadkat, Shyamal and others},
  journal={arXiv preprint arXiv:2303.08774},
  year={2023}
}

@article{touvron2023llama,
  title={Llama: Open and efficient foundation language models},
  author={Touvron, Hugo and Lavril, Thibaut and Izacard, Gautier and Martinet, Xavier and Lachaux, Marie-Anne and Lacroix, Timoth{\'e}e and Rozi{\`e}re, Baptiste and Goyal, Naman and Hambro, Eric and Azhar, Faisal and others},
  journal={arXiv preprint arXiv:2302.13971},
  year={2023}
}

@article{comanici2025gemini,
  title={Gemini 2.5: Pushing the frontier with advanced reasoning, multimodality, long context, and next generation agentic capabilities},
  author={Comanici, Gheorghe and Bieber, Eric and Schaekermann, Mike and Pasupat, Ice and Sachdeva, Noveen and Dhillon, Inderjit and Blistein, Marcel and Ram, Ori and Zhang, Dan and Rosen, Evan and others},
  journal={arXiv preprint arXiv:2507.06261},
  year={2025}
}

@article{team2026kimi,
  title={Kimi k3: Open frontier intelligence},
  author={Team, Kimi and Bai, Tongtong and Bai, Yifan and Bao, Yiping and Cai, Jianfeng and Cai, Xinyuan and Cao, Peizhou and Cao, Yuxuan and Chai, Ziwei and Charles, Y and others},
  journal={arXiv preprint arXiv:2607.24653},
  year={2026}
}

@article{team2026qwen3,
  title={Qwen3. 5-omni technical report},
  author={Team, Qwen},
  journal={arXiv preprint arXiv:2604.15804},
  year={2026}
}

@inproceedings{kwon2023efficient,
  title={Efficient memory management for large language model serving with pagedattention},
  author={Kwon, Woosuk and Li, Zhuohan and Zhuang, Siyuan and Sheng, Ying and Zheng, Lianmin and Yu, Cody Hao and Gonzalez, Joseph and Zhang, Hao and Stoica, Ion},
  booktitle={Proceedings of the 29th symposium on operating systems principles},
  pages={611--626},
  year={2023}
}

@article{zheng2024sglang,
  title={Sglang: Efficient execution of structured language model programs},
  author={Zheng, Lianmin and Yin, Liangsheng and Xie, Zhiqiang and Sun, Chuyue and Huang, Jeff and Yu, Cody H and Cao, Shiyi and Kozyrakis, Christos and Stoica, Ion and Gonzalez, Joseph E and others},
  journal={Advances in neural information processing systems},
  volume={37},
  pages={62557--62583},
  year={2024}
}

@article{tan2026rtp,
  title={RTP-LLM: High-Performance Alibaba LLM Inference Engine},
  author={Tan, Boyu and Guo, Jiarui and Lv, Zongwei and Sun, Hanbo and Yang, Tong and Liu, Kan and Shi, Xinfei and Hu, Zetao and Yu, Yaxin and Zhang, Chi and others},
  journal={arXiv preprint arXiv:2605.29639},
  year={2026}
}

@inproceedings{li2023sequence,
  title={Sequence parallelism: Long sequence training from system perspective},
  author={Li, Shenggui and Xue, Fuzhao and Baranwal, Chaitanya and Li, Yongbin and You, Yang},
  booktitle={Proceedings of the 61st Annual Meeting of the Association for Computational Linguistics (Volume 1: Long Papers)},
  pages={2391--2404},
  year={2023}
}

@inproceedings{liu2024ringattention,
  title={Ringattention with blockwise transformers for near-infinite context},
  author={Liu, Hao and Zaharia, Matei and Abbeel, Pieter},
  booktitle={International Conference on Learning Representations},
  volume={2024},
  pages={3992--4008},
  year={2024}
}

@article{brandon2023striped,
  title={Striped attention: Faster ring attention for causal transformers},
  author={Brandon, William and Nrusimha, Aniruddha and Qian, Kevin and Ankner, Zachary and Jin, Tian and Song, Zhiye and Ragan-Kelley, Jonathan},
  journal={arXiv preprint arXiv:2311.09431},
  year={2023}
}

@article{yang2025context,
  title={Context parallelism for scalable million-token inference},
  author={Yang, Amy and Yang, Jingyi and Ibrahim, Aya and Xie, Xinfeng and Tang, Bangsheng and Sizov, Grigory and Park, Jongsoo and Huang, Jianyu},
  journal={Proceedings of Machine Learning and Systems},
  volume={7},
  year={2025}
}

@inproceedings{jiang2025dcp,
  title={DCP: Addressing Input Dynamism In Long-Context Training via Dynamic Context Parallelism},
  author={Jiang, Chenyu and Cai, Zhenkun and Tian, Ye and Jia, Zhen and Wang, Yida and Wu, Chuan},
  booktitle={Proceedings of the ACM SIGOPS 31st Symposium on Operating Systems Principles},
  pages={221--236},
  year={2025}
}

@article{jacobs2023deepspeed,
  title={Deepspeed ulysses: System optimizations for enabling training of extreme long sequence transformer models},
  author={Jacobs, Sam Ade and Tanaka, Masahiro and Zhang, Chengming and Zhang, Minjia and Song, Shuaiwen Leon and Rajbhandari, Samyam and He, Yuxiong},
  journal={arXiv preprint arXiv:2309.14509},
  year={2023}
}

@article{fang2024usp,
  title={Usp: A unified sequence parallelism approach for long context generative ai},
  author={Fang, Jiarui and Zhao, Shangchun},
  journal={arXiv preprint arXiv:2405.07719},
  year={2024}
}

@article{gu2024loongtrain,
  title={Loongtrain: Efficient training of long-sequence llms with head-context parallelism},
  author={Gu, Diandian and Sun, Peng and Hu, Qinghao and Huang, Ting and Chen, Xun and Xiong, Yingtong and Wang, Guoteng and Chen, Qiaoling and Zhao, Shangchun and Fang, Jiarui and others},
  journal={arXiv preprint arXiv:2406.18485},
  year={2024}
}

@inproceedings{wang2025flexsp,
  title={Flexsp: Accelerating large language model training via flexible sequence parallelism},
  author={Wang, Yujie and Wang, Shiju and Zhu, Shenhan and Fu, Fangcheng and Liu, Xinyi and Xiao, Xuefeng and Li, Huixia and Li, Jiashi and Wu, Faming and Cui, Bin},
  booktitle={Proceedings of the 30th ACM International Conference on Architectural Support for Programming Languages and Operating Systems, Volume 2},
  pages={421--436},
  year={2025}
}

@inproceedings{ge2025bytescale,
  title={ByteScale: Communication-Efficient Scaling of LLM Training with a 2048K Context Length on 16384 GPUs},
  author={Ge, Hao and Feng, Junda and Huang, Qi and Fu, Fangcheng and Nie, Xiaonan and Zuo, Lei and Lin, Haibin and Cui, Bin and Liu, Xin},
  booktitle={Proceedings of the ACM SIGCOMM 2025 Conference},
  pages={963--978},
  year={2025}
}

@article{lin2024infinite,
  title={Infinite-llm: Efficient llm service for long context with distattention and distributed kvcache},
  author={Lin, Bin and Zhang, Chen and Peng, Tao and Zhao, Hanyu and Xiao, Wencong and Sun, Minmin and Liu, Anmin and Zhang, Zhipeng and Li, Lanbo and Qiu, Xiafei and others},
  journal={arXiv preprint arXiv:2401.02669},
  year={2024}
}

@inproceedings{zhong2024distserve,
  title={$\{$DistServe$\}$: Disaggregating prefill and decoding for goodput-optimized large language model serving},
  author={Zhong, Yinmin and Liu, Shengyu and Chen, Junda and Hu, Jianbo and Zhu, Yibo and Liu, Xuanzhe and Jin, Xin and Zhang, Hao},
  booktitle={18th USENIX Symposium on Operating Systems Design and Implementation (OSDI 24)},
  pages={193--210},
  year={2024}
}

@article{qin2024mooncake,
  title={Mooncake: A kvcache-centric disaggregated architecture for llm serving},
  author={Qin, Ruoyu and Li, Zheming and He, Weiran and Cui, Jialei and Tang, Heyi and Ren, Feng and Ma, Teng and Cai, Shangming and Zhang, Yineng and Zhang, Mingxing and others},
  journal={ACM Transactions on Storage},
  year={2024},
  publisher={ACM New York, NY}
}

@article{shi2024keep,
  title={Keep the cost down: A review on methods to optimize LLM's KV-cache consumption},
  author={Shi, Luohe and Zhang, Hongyi and Yao, Yao and Li, Zuchao and Zhao, Hai},
  journal={arXiv preprint arXiv:2407.18003},
  year={2024}
}

@inproceedings{wang2025kvcache,
  title={$\{$KVCache$\}$ Cache in the Wild: Characterizing and Optimizing $\{$KVCache$\}$ Cache at a Large Cloud Provider},
  author={Wang, Jiahao and Han, Jinbo and Wei, Xingda and Shen, Sijie and Zhang, Dingyan and Fang, Chenguang and Chen, Rong and Yu, Wenyuan and Chen, Haibo},
  booktitle={2025 USENIX Annual Technical Conference (USENIX ATC 25)},
  pages={465--482},
  year={2025}
}

@article{wang2026prefixkv,
  title={Prefixkv: Adaptive prefix kv cache is what vision instruction-following models need for efficient generation},
  author={Wang, Ao and Chen, Hui and Tan, Jianchao and Zhang, Kefeng and Cai, Xunliang and Lin, Zijia and Han, Jungong and others},
  journal={Advances in Neural Information Processing Systems},
  volume={38},
  pages={94456--94482},
  year={2026}
}

@inproceedings{ye2024chunkattention,
  title={Chunkattention: Efficient self-attention with prefix-aware kv cache and two-phase partition},
  author={Ye, Lu and Tao, Ze and Huang, Yong and Li, Yang},
  booktitle={Proceedings of the 62nd Annual Meeting of the Association for Computational Linguistics (Volume 1: Long Papers)},
  pages={11608--11620},
  year={2024}
}

@article{pan2026kvflow,
  title={KVFlow: Efficient prefix caching for accelerating LLM-based multi-agent workflows},
  author={Pan, Zaifeng and Patel, Ajjkumar and Shen, Yipeng and Hu, Zhengding and Guan, Yue and Li, Wan-Lu and Qin, Lianhui and Wang, Yida and Ding, Yufei},
  journal={Advances in Neural Information Processing Systems},
  volume={38},
  pages={126246--126265},
  year={2026}
}

@article{yang2026learned,
  title={Learned prefix caching for efficient LLM inference},
  author={Yang, Dongsheng and Li, Austin and Li, Kai and Lloyd, Wyatt},
  journal={Advances in Neural Information Processing Systems},
  volume={38},
  pages={45770--45792},
  year={2026}
}

@article{li2025hotprefix,
  title={Hotprefix: Hotness-aware kv cache scheduling for efficient prefix sharing in llm inference systems},
  author={Li, Yuhang and Gu, Rong and Huan, Chengying and Wang, Zhibin and Yao, Renjie and Tian, Chen and Chen, Guihai},
  journal={Proceedings of the ACM on Management of Data},
  volume={3},
  number={4},
  pages={1--27},
  year={2025},
  publisher={ACM New York, NY, USA}
}

@article{gim2023prompt,
  title={Prompt cache: Modular attention reuse for low-latency inference},
  author={Gim, In and Chen, Guojun and Lee, Seung-seob and Sarda, Nikhil and Khandelwal, Anurag and Zhong, Lin},
  journal={arXiv preprint arXiv:2311.04934},
  year={2023}
}

@inproceedings{srivatsa2025preble,
  title={Preble: Efficient distributed prompt scheduling for llm serving},
  author={Srivatsa, Vikranth and He, Zijian and Abhyankar, Reyna and Li, Dongming and Zhang, Yiying},
  booktitle={International conference on learning representations},
  volume={2025},
  pages={37057--37082},
  year={2025}
}

@article{zhang2025pqcache,
  title={Pqcache: Product quantization-based kvcache for long context llm inference},
  author={Zhang, Hailin and Ji, Xiaodong and Chen, Yilin and Fu, Fangcheng and Miao, Xupeng and Nie, Xiaonan and Chen, Weipeng and Cui, Bin},
  journal={Proceedings of the ACM on Management of Data},
  volume={3},
  number={3},
  pages={1--30},
  year={2025},
  publisher={ACM New York, NY, USA}
}

@inproceedings{wu2024loongserve,
  title={Loongserve: Efficiently serving long-context large language models with elastic sequence parallelism},
  author={Wu, Bingyang and Liu, Shengyu and Zhong, Yinmin and Sun, Peng and Liu, Xuanzhe and Jin, Xin},
  booktitle={Proceedings of the ACM SIGOPS 30th Symposium on Operating Systems Principles},
  pages={640--654},
  year={2024}
}

@article{zhang2026simple,
  title={Simple is Better: Multiplication May Be All You Need for LLM Request Scheduling},
  author={Zhang, Dingyan and Han, Jinbo and Zhang, Kaixi and Wei, Xingda and Shen, Sijie and Fang, Chenguang and Yu, Wenyuan and Zhou, Jingren and Chen, Rong},
  journal={arXiv preprint arXiv:2603.15202},
  year={2026}
}

@inproceedings{yu2022orca,
  title={Orca: A distributed serving system for $\{$Transformer-Based$\}$ generative models},
  author={Yu, Gyeong-In and Jeong, Joo Seong and Kim, Geon-Woo and Kim, Soojeong and Chun, Byung-Gon},
  booktitle={16th USENIX symposium on operating systems design and implementation (OSDI 22)},
  pages={521--538},
  year={2022}
}

@inproceedings{sun2024llumnix,
  title={Llumnix: Dynamic scheduling for large language model serving},
  author={Sun, Biao and Huang, Ziming and Zhao, Hanyu and Xiao, Wencong and Zhang, Xinyi and Li, Yong and Lin, Wei},
  booktitle={18th USENIX symposium on operating systems design and implementation (OSDI 24)},
  pages={173--191},
  year={2024}
}

@article{chen2026nanocp,
  title={NanoCP: Request-Level Dynamic Context Parallelism for Data-Expert Parallel Decoding},
  author={Chen, Jiefei and Lin, Binbin and Ma, Jinming and Duan, Jiangfei and Duanmu, Haojie and Liu, Hao and Cheng, Qinxiu and Li, Xiuhong and Pei, Zhilin and Wang, Hui and others},
  journal={arXiv preprint arXiv:2605.21100},
  year={2026}
}

@article{wu2025tokenlake,
  title={TokenLake: A Unified Segment-level Prefix Cache Pool for Fine-grained Elastic Long-Context LLM Serving},
  author={Wu, Bingyang and Zhang, Zili and Zhong, Yinmin and Huang, Guanzhe and Zhu, Yibo and Liu, Xuanzhe and Jin, Xin},
  journal={arXiv preprint arXiv:2508.17219},
  year={2025}
}

@article{sun2026hygen,
  title={Hygen: Efficient llm serving via elastic online-offline request co-location},
  author={Sun, Ting and Wang, Penghan and Lai, Fan},
  journal={Advances in Neural Information Processing Systems},
  volume={38},
  pages={14910--14937},
  year={2026}
}

@article{jiang2025neo,
  title={Neo: Saving gpu memory crisis with cpu offloading for online llm inference},
  author={Jiang, Xuanlin and Zhou, Yang and Cao, Shiyi and Stoica, Ion and Yu, Minlan},
  journal={Proceedings of Machine Learning and Systems},
  volume={7},
  year={2025}
}

@inproceedings{agrawal2024taming,
  title={Taming $\{$Throughput-Latency$\}$ tradeoff in $\{$LLM$\}$ inference with $\{$Sarathi-Serve$\}$},
  author={Agrawal, Amey and Kedia, Nitin and Panwar, Ashish and Mohan, Jayashree and Kwatra, Nipun and Gulavani, Bhargav and Tumanov, Alexey and Ramjee, Ramachandran},
  booktitle={18th USENIX symposium on operating systems design and implementation (OSDI 24)},
  pages={117--134},
  year={2024}
}

@inproceedings{wu2026fastserve,
  title={$\{$FastServe$\}$:$\{$Iteration-Level$\}$ Preemptive Scheduling for Large Language Model Inference},
  author={Wu, Bingyang and Zhong, Yinmin and Zhang, Zili and Liu, Shengyu and Liu, Fangyue and Sun, Yuanhang and Huang, Gang and Liu, Xuanzhe and Jin, Xin},
  booktitle={23rd USENIX Symposium on Networked Systems Design and Implementation (NSDI 26)},
  pages={57--74},
  year={2026}
}

@article{duan2024muxserve,
  title={MuxServe: flexible spatial-temporal multiplexing for multiple LLM serving},
  author={Duan, Jiangfei and Lu, Runyu and Duanmu, Haojie and Li, Xiuhong and Zhang, Xingcheng and Lin, Dahua and Stoica, Ion and Zhang, Hao},
  journal={arXiv preprint arXiv:2404.02015},
  year={2024}
}

@inproceedings{li2023alpaserve,
  title={$\{$AlpaServe$\}$: Statistical multiplexing with model parallelism for deep learning serving},
  author={Li, Zhuohan and Zheng, Lianmin and Zhong, Yinmin and Liu, Vincent and Sheng, Ying and Jin, Xin and Huang, Yanping and Chen, Zhifeng and Zhang, Hao and Gonzalez, Joseph E and others},
  booktitle={17th USENIX Symposium on Operating Systems Design and Implementation (OSDI 23)},
  pages={663--679},
  year={2023}
}

@inproceedings{patel2024splitwise,
  title={Splitwise: Efficient generative llm inference using phase splitting},
  author={Patel, Pratyush and Choukse, Esha and Zhang, Chaojie and Shah, Aashaka and Goiri, {\'I}{\~n}igo and Maleki, Saeed and Bianchini, Ricardo},
  booktitle={2024 ACM/IEEE 51st Annual International Symposium on Computer Architecture (ISCA)},
  pages={118--132},
  year={2024},
  organization={IEEE}
}

@article{hu2024inference,
  title={Inference without interference: Disaggregate llm inference for mixed downstream workloads},
  author={Hu, Cunchen and Huang, Heyang and Xu, Liangliang and Chen, Xusheng and Xu, Jiang and Chen, Shuang and Feng, Hao and Wang, Chenxi and Wang, Sa and Bao, Yungang and others},
  journal={arXiv preprint arXiv:2401.11181},
  year={2024}
}

@article{hu2024memserve,
  title={Memserve: Context caching for disaggregated llm serving with elastic memory pool},
  author={Hu, Cunchen and Huang, Heyang and Hu, Junhao and Xu, Jiang and Chen, Xusheng and Xie, Tao and Wang, Chenxi and Wang, Sa and Bao, Yungang and Sun, Ninghui and others},
  journal={arXiv preprint arXiv:2406.17565},
  year={2024}
}

@article{liu2025lmcache,
  title={Lmcache: An efficient KV cache layer for enterprise-scale LLM inference},
  author={Liu, Yuhan and Cheng, Yihua and Yao, Jiayi and An, Yuwei and Chen, Xiaokun and Feng, Shaoting and Huang, Yuyang and Shen, Samuel and Zhang, Rui and Du, Kuntai and others},
  journal={arXiv preprint arXiv:2510.09665},
  year={2025}
}

@article{mo2026serving,
  title={Serving Hybrid LLM Loads with SLO Guarantees Using CPU-GPU Attention Piggybacking},
  author={Mo, Zizhao and Chen, Junlin and Xu, Huanle and Xu, Chengzhong},
  journal={Proceedings of the ACM on Management of Data},
  volume={4},
  number={3 (SIGMOD},
  pages={1--26},
  year={2026},
  publisher={ACM New York, NY, USA}
}

@article{gao2025apt,
  title={Apt-serve: Adaptive request scheduling on hybrid cache for scalable llm inference serving},
  author={Gao, Shihong and Zhang, Xin and Shen, Yanyan and Chen, Lei},
  journal={Proceedings of the ACM on Management of Data},
  volume={3},
  number={3},
  pages={1--28},
  year={2025},
  publisher={ACM New York, NY, USA}
}

@article{bai2026alignedserve,
  title={AlignedServe: Orchestrating Prefix-aware Batching to Build a High-throughput and Computing-efficient LLM Serving System},
  author={Bai, Fengyao and Zhang, Hongbin and Chen, Zhitao and Du, Jiangsu and Chen, Zhiguang and Lu, Yutong},
  journal={Proceedings of the ACM on Management of Data},
  volume={4},
  number={3 (SIGMOD},
  pages={1--25},
  year={2026},
  publisher={ACM New York, NY, USA}
}

@inproceedings{miao2024spotserve,
  title={Spotserve: Serving generative large language models on preemptible instances},
  author={Miao, Xupeng and Shi, Chunan and Duan, Jiangfei and Xi, Xiaoli and Lin, Dahua and Cui, Bin and Jia, Zhihao},
  booktitle={Proceedings of the 29th ACM International Conference on Architectural Support for Programming Languages and Operating Systems, Volume 2},
  pages={1112--1127},
  year={2024}
}

@article{bai2026pipelive,
  title={PipeLive: Efficient Live In-place Pipeline Parallelism Reconfiguration for Dynamic LLM Serving},
  author={Bai, Xu and Islam, Muhammed Tawfiqul and Wang, Chen and Toosi, Adel N},
  journal={arXiv preprint arXiv:2604.12171},
  year={2026}
}

@article{yuan2026remp,
  title={ReMP: Low-Downtime Runtime Model-Parallelism Reconfiguration for LLM Serving},
  author={Yuan, Haipeng and Zheng, Kaining and Bai, Yongshu and Zhang, Yuchen and Zhang, Yunquan and Wu, Baodong and Gao, Xiang and Cheng, Daning},
  journal={arXiv preprint arXiv:2606.18741},
  year={2026}
}

@inproceedings{gao2026flying,
  title={FLYING SERVING: On-the-Fly Parallelism Switching for Large Language Model Serving},
  author={Gao, Shouwei and Yin, Junqi and Wang, Feiyi and Dong, Wenqian},
  booktitle={Proceedings of the 40th ACM International Conference on Supercomputing},
  pages={17--29},
  year={2026}
}

@article{lin2026kvdrive,
  title={KVDrive: A Holistic Multi-Tier KV Cache Management System for Long-Context LLM Inference},
  author={Lin, Jian and Mi, Jiazhi and Hong, Zicong and Wang, Haodong and Liu, Qianli and Zhang, Haoyue and Li, Peng and Guo, Song},
  journal={Proceedings of the ACM on Management of Data},
  volume={4},
  number={3 (SIGMOD},
  pages={1--25},
  year={2026},
  publisher={ACM New York, NY, USA}
}

@article{jiang2024chameleon,
  title={Chameleon: A Heterogeneous and Disaggregated Accelerator System for Retrieval-Augmented Language Models},
  author={Jiang, Wenqi and Zeller, Marco and Waleffe, Roger and Hoefler, Torsten and Alonso, Gustavo},
  journal={Proceedings of the VLDB Endowment},
  volume={18},
  number={1},
  pages={42--52},
  year={2024},
  publisher={VLDB Endowment}
}

@article{yang2026beluga,
  title={Beluga: A cxl-based memory architecture for scalable and efficient llm kvcache management},
  author={Yang, Xinjun and Hu, Qingda and Li, Junru and Li, Feifei and Zhu, Yicong and Zhou, Yuqi and Lin, Qiuru and Dai, Jian and Kong, Yang and Zhang, Jiayu and others},
  journal={Proceedings of the ACM on Management of Data},
  volume={4},
  number={1 (SIGMOD},
  pages={1--29},
  year={2026},
  publisher={ACM New York, NY, USA}
}

@article{hong2025sola,
  title={Sola: Optimizing slo attainment for large language model serving with state-aware scheduling},
  author={Hong, Ke and Li, Xiuhong and Chen, Lufang and Mao, Qiuli and Dai, Guohao and Ning, Xuefei and Yan, Shengen and Liang, Yun and Wang, Yu},
  journal={Proceedings of Machine Learning and Systems},
  volume={7},
  year={2025}
}

@inproceedings{gong2025past,
  title={Past-future scheduler for llm serving under sla guarantees},
  author={Gong, Ruihao and Bai, Shihao and Wu, Siyu and Fan, Yunqian and Wang, Zaijun and Li, Xiuhong and Yang, Hailong and Liu, Xianglong},
  booktitle={Proceedings of the 30th ACM International Conference on Architectural Support for Programming Languages and Operating Systems, Volume 2},
  pages={798--813},
  year={2025}
}

@article{yang2025lserve,
  title={Lserve: Efficient long-sequence llm serving with unified sparse attention},
  author={Yang, Shang and Guo, Junxian and Tang, Haotian and Hu, Qinghao and Xiao, Guangxuan and Tang, Jiaming and Lin, Yujun and Liu, Zhijian and Lu, Yao and Han, Song},
  journal={Proceedings of Machine Learning and Systems},
  volume={7},
  year={2025}
}

@article{chen2025gyges,
  title={Gyges: Dynamic Cross-Instance Parallelism Transformation for Efficient LLM Inference},
  author={Chen, Haoyu and Li, Xue and Qian, Kun and Guan, Yu and Zhao, Jin and Wang, Xin},
  journal={arXiv preprint arXiv:2509.19729},
  year={2025}
}

@article{wu2025unlock,
  title={Unlock the Potential of Fine-grained LLM Serving via Dynamic Module Scaling},
  author={Wu, Jingfeng and He, Yiyuan and Xu, Minxian and Gao, Xitong and Ye, Kejiang and Xu, Chengzhong},
  journal={arXiv preprint arXiv:2507.18006},
  year={2025}
}

@article{jiang2026oserve,
  title={Oserve: Accelerating llm serving via spatial-temporal workload orchestration},
  author={Jiang, Youhe and Fu, Fangcheng and Wang, Taiyi and He, Guoliang and Yoneki, Eiko},
  journal={arXiv preprint arXiv:2602.12151},
  year={2026}
}

@article{agrawal2024medha,
  title={Medha: Efficiently serving multi-million context length LLM inference requests without approximations},
  author={Agrawal, Amey and Qiu, Haoran and Chen, Junda and Goiri, {\'I}{\~n}igo and Zhang, Chaojie and Shahid, Rayyan and Ramjee, Ramachandran and Tumanov, Alexey and Choukse, Esha},
  journal={arXiv preprint arXiv:2409.17264},
  year={2024}
}

@article{lewis2020retrieval,
  title={Retrieval-augmented generation for knowledge-intensive nlp tasks},
  author={Lewis, Patrick and Perez, Ethan and Piktus, Aleksandra and Petroni, Fabio and Karpukhin, Vladimir and Goyal, Naman and K{\"u}ttler, Heinrich and Lewis, Mike and Yih, Wen-tau and Rockt{\"a}schel, Tim and others},
  journal={Advances in neural information processing systems},
  volume={33},
  pages={9459--9474},
  year={2020}
}

@inproceedings{bai2025longbench,
  title={Longbench v2: Towards deeper understanding and reasoning on realistic long-context multitasks},
  author={Bai, Yushi and Tu, Shangqing and Zhang, Jiajie and Peng, Hao and Wang, Xiaozhi and Lv, Xin and Cao, Shulin and Xu, Jiazheng and Hou, Lei and Dong, Yuxiao and others},
  booktitle={Proceedings of the 63rd Annual Meeting of the Association for Computational Linguistics (Volume 1: Long Papers)},
  pages={3639--3664},
  year={2025}
}

@article{talebirad2023multi,
  title={Multi-agent collaboration: Harnessing the power of intelligent llm agents},
  author={Talebirad, Yashar and Nadiri, Amirhossein},
  journal={arXiv preprint arXiv:2306.03314},
  year={2023}
}

@article{vaswani2017attention,
  title={Attention is all you need},
  author={Vaswani, Ashish and Shazeer, Noam and Parmar, Niki and Uszkoreit, Jakob and Jones, Llion and Gomez, Aidan N and Kaiser, {\L}ukasz and Polosukhin, Illia},
  journal={Advances in neural information processing systems},
  volume={30},
  year={2017}
}

@inproceedings{wang2025burstgpt,
  title={Burstgpt: A real-world workload dataset to optimize llm serving systems},
  author={Wang, Yuxin and Chen, Yuhan and Li, Zeyu and Kang, Xueze and Fang, Yuchu and Zhou, Yeju and Zheng, Yang and Tang, Zhenheng and He, Xin and Guo, Rui and others},
  booktitle={Proceedings of the 31st ACM SIGKDD Conference on Knowledge Discovery and Data Mining V. 2},
  pages={5831--5841},
  year={2025}
}

@article{wang2025llm,
  title={Llm serving optimization with variable prefill and decode lengths},
  author={Wang, Meixuan and Ye, Yinyu and Zhou, Zijie},
  journal={arXiv preprint arXiv:2508.06133},
  year={2025}
}

@misc{toolagent, 
  title={Mooncake trace}, 
  howpublished={\url{https://github.com/kvcacheai/Mooncake/blob/main/FAST25-release/traces/toolagent_trace.jsonl}}
}

@inproceedings{an2024eval,
  title={L-eval: Instituting standardized evaluation for long context language models},
  author={An, Chenxin and Gong, Shansan and Zhong, Ming and Zhao, Xingjian and Li, Mukai and Zhang, Jun and Kong, Lingpeng and Qiu, Xipeng},
  booktitle={Proceedings of the 62nd Annual Meeting of the Association for Computational Linguistics (Volume 1: Long Papers)},
  pages={14388--14411},
  year={2024}
}
}

\end{document}